# Grain boundary interstitial segregation in substitutional binary alloys

Zuoyong Zhang and Chuang Deng*

*Department of Mechanical Engineering, University of Manitoba, Winnipeg, Manitoba R3T 5V6, Canada*

* Corresponding author: Chuang.Deng@umanitoba.ca

**Graphical abstract**

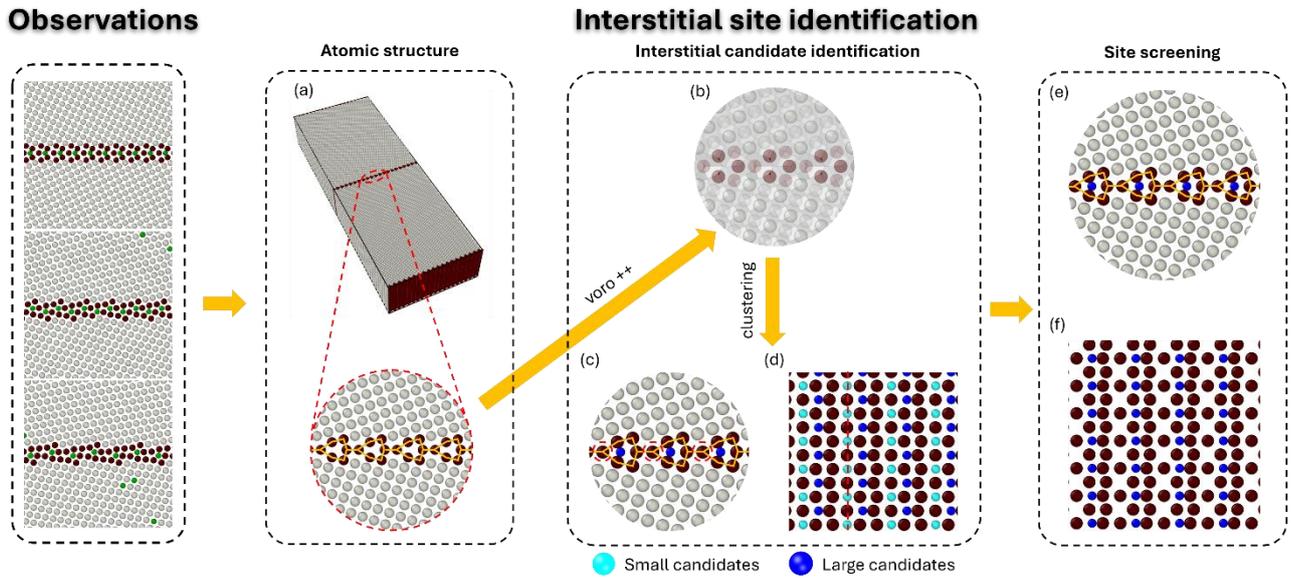

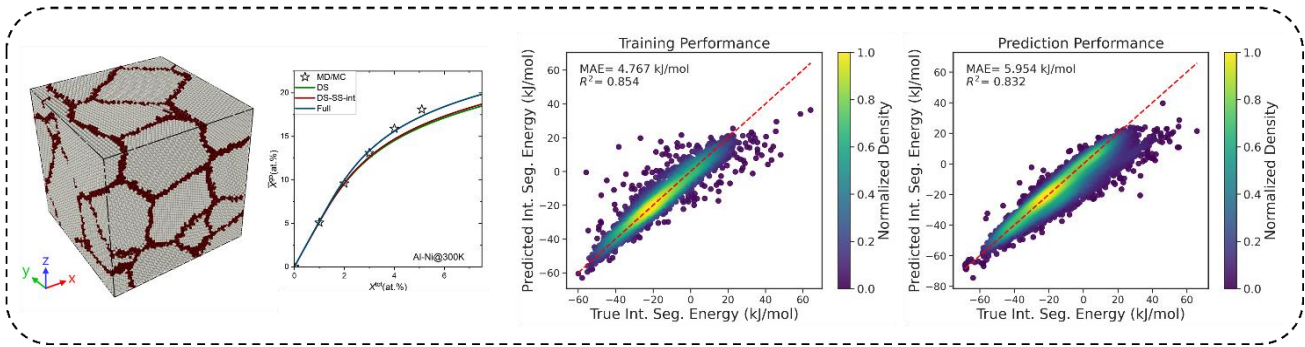

**Abstract**

Grain boundary (GB) segregation is a powerful approach for optimizing the thermal and mechanical properties of metal alloys. In this study, we report significant GB interstitial segregation in a representative substitutional binary alloy system (Al-Ni) through atomistic simulations, challenging prevailing assumptions in the literature. Our findings show that Ni atoms preferentially segregate to interstitial sites within numerous kite-like GB structures in the Al bicrystals. An intriguing interplanar interstitial segregation pattern was also observed and analyzed. Additionally, interstitial segregation can induce unexpected GB transitions, such as kite transitions and nano-faceting, due to the existence of small interstitial sites. Building upon these observations, we developed a robust method to systematically identify the interstitial candidate sites for accommodating solutes at GBs. This approach combines site detection with structural filtering to produce distributions of interstitial sites that closely match atomistic simulation results. Applied to nanocrystalline alloys, this method enabled the calculation of interstitial segregation energies, significantly improving GB segregation predictions for the Al-Ni system. Furthermore, machine learning models using smooth overlap of atomic positions descriptors successfully predicted per-site interstitial segregation energy. This study highlights the critical role of GB interstitial segregation in advancing our understanding of solute behavior and provides valuable insights for designing next-generation alloys.

**Keywords:**

Grain boundary interstitial segregation; Atomistic simulations; Grain boundary structural transition.

**1 Introduction**

Pure metals are rarely used in engineering applications due to their relatively low strength and poor thermal stability, especially in the presence of ultrafine grain structures. These fine-grained structures exhibit high grain boundary (GB) excess energy, which compromises their thermal stability and leads to grain growth even at modest temperatures [1–6]. This grain coarsening can significantly degrade their mechanical performance, as described by the Hall-Petch relationship [7–9]. Thus, it is crucial to develop strategies to mitigate this effect and maintain the stability of the GB network.

Over the past decades, solute segregation at GBs has emerged as a promising strategy for alloy design [10–15]. Experimental studies [16–18] have shown that GB segregation can stabilize GB network and effectively inhibit grain growth, even under highly aggressive conditions [19]. This stabilization is attributed to the formation of thermodynamically equilibrated GB network, achieved by reducing GB excess energy, the primary driving force for grain growth, through decorating them with solute atoms [20–22]. GB segregation has also been reported as an effective method to inhibit or delay the transition of Hall-Petch to inverse Hall-Petch behaviors by stabilizing the ultrafine-grained structures [23–25]. Furthermore, it can alter GB complexions [26–28], structures [29–32], and topologies [33], and even deformation mechanisms [34–36]. These effects enable the development of materials with exceptional properties, including ultra-high strength [37,38], strength-ductility synergy [39–42], and enhanced fatigue resistance [43,44], which are quite challenging to achieve through conventional metallurgical methods. These findings highlight GB segregation as a powerful tool for improving the mechanical properties of materials. However, the accumulation of dopants through GB segregation can also lead to some catastrophic effects, such as embrittlement [45–48], decohesion of GB structures [49–51], and increased susceptibility to intergranular corrosion [52,53]. Therefore, it is essential to carefully select appropriate solutes and optimize treatment conditions for successful applications of GB segregation engineering.

Recently, a spectral approach [54–56] has been developed for predicting GB segregation behaviors. It has been extensively used to investigate the effects of various factors on the GB segregation, including grain size [57], triple junctions [58], entropy [59–61], hydrostatic pressure [62], and crystallographic characters [63]. Incorporating solute-solute interactions into spectral models [64–67] has further enhanced the accuracy of GB segregation predictions.

However, most previous studies focusing on substitutional alloys assume that segregated solute atoms, similar to those in the grain interior, exclusively occupy substitutional sites within GBs. This leaves a fundamental question unanswered: can interstitial GB segregation also occur in substitutional alloys? Recent first-principles studies [68–71] have demonstrated that Cu atoms preferentially segregate to hollow interstitial sites within the Al symmetric tilt grain boundaries (STGBs), driven by their lower segregation energy than substitutional sites. This phenomenon has also been confirmed by experimental observations [72]. These findings highlight the potential for GB interstitial segregation in typical substitutional alloy systems, offering new insights into solute segregation behaviors and extending the understanding of segregation mechanisms beyond traditional substitutional behaviors.

When discussing interstitial segregation, the solutes typically refer to those small nonmetallic species, such as C, N, O, and B, etc. Their small atomic sizes allow them to preferentially occupy interstitial sites within metals, making them ideal candidates for interstitial segregation. However, GB interstitial segregation in nanocrystalline (NC) substitutional alloys remains an area that has not been systematically explored. Additionally, the identification of interstitial candidate sites in these alloys is yet to be addressed. In this study, we employ hybrid molecular dynamics (MD)/Monte Carlo (MC) simulations to capture the interstitial segregation behavior of Ni in Al, a typical substitutional alloy system, using both the 174 Al coincidence site lattice (CSL) GBs [73] and general NC models. The results reveal that interstitial segregation of Ni can commonly be observed within numerous CSL GBs in Al bicrystals. Building upon these findings, this work seeks to address the following two critical questions:

(1) How can we identify the potential interstitial sites in both bicrystal and NC substitutional alloys?
(2) How does the interstitial segregation affect the GB segregation predictions in typical NC substitutional alloys?

Through this study, we aim to deepen the understanding of GB interstitial segregation and its implications for alloy design.

## 2 Method

### *2.1 Hybrid MD/MC simulations*

In this study, we selected the Al-Ni system as a representative case to investigate the interstitial segregation behaviors in substitutional alloys. A key reason for this choice is the slight size disparity between the face-centered cubic (FCC) Ni and Al atoms. We hypothesize that this size difference plays a critical role in facilitating the interstitial segregation of Ni in Al, similar to the previously reported preference of Cu to segregate to hollow interstitial sites in Al STGBs [68,71,72]. Another reason for selecting this system is the availability of extensive data in the literature [55,56,66], which allows for meaningful comparisons.

All atomistic simulations were performed using the open-source software package LAMMPS [74]. Periodic boundary conditions were applied in all directions of the simulation box. Visualization and structural analysis were performed using OVITO [75], and the additive-common neighbor analysis (a-CNA) method [76] was employed to identify atomic structures. Interatomic interactions were modeled using the embedded-atom method (EAM) potential for the Al-Ni system developed by Purja Pun and Mishin [77], which has been widely used in studies of GB segregation [55,62,66,67]. The 174 Al bicrystals were constructed by replicating the original CSL GB models from Ref. [73] between 2 and 16 times in the $x$ and $z$ directions, depending on the size of the initial GB models, as shown in Fig. 1(a). The dataset features Al CSL GBs with $\Sigma 3$, 5, 9, 11, and 13, as well as (001), (110), and (111) STGBs. Each bicrystal model was optimized using a conjugate gradient (CG) algorithm to prepare for the subsequent hybrid MD/MC simulations. Hybrid MD/MC simulations were then performed using the variance-constrained semi-grand-canonical (VC-SGC) ensemble [78] at 300 K with all simulation details described in our previous work [66].

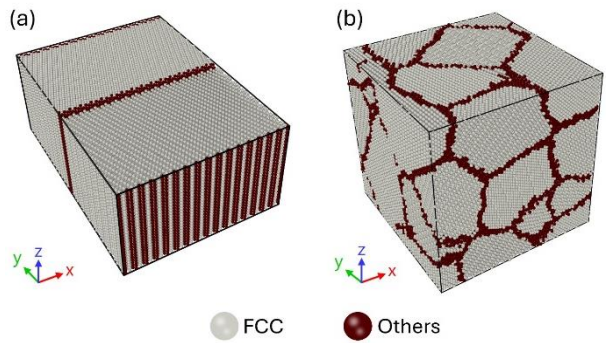

**Fig. 1** (a) A representative of the optimized Al bicrystal models. (b) The fully relaxed Al NC model has 10 randomly oriented grains with an average grain size of 9.1 nm.

### *2.2 Per-site interstitial segregation energy*

A fully relaxed NC Al specimen with the size of 16³ *nm*³ (referred to as Al-16) and containing 10 randomly oriented grains, as shown in Fig. 1(b), was used to determine the per-site interstitial segregation energy. The Al-16 model was provided by Wagih and Schuh [54] from their previous work and has also been used in other studies [62,66,79,80]. Before assessing the interstitial segregation energy, it is crucial to identify the interstitial candidate sites within the GB network. The method for interstitial site identification is detailed in Section 4.

The per-site interstitial segregation energy of an interstitial Ni atom is defined as the difference in system energy between a solute atom occupying a GB interstitial site and a bulk substitutional site, minus the reference energy of a solvent atom:

$$\Delta E_i^{\text{SS,int}} = E_i^{\text{Ni,int}} - E_{\text{ref}}^{\text{Ni,bulk}} - \bar{E}_{\text{ref}}^{\text{Al,bulk}}. \tag{1}$$

where $\Delta E_i^{\text{SS,int}}$ is the per-site interstitial segregation energy for inserting a solute atom into an interstitial site *i*. The superscripts SS and int denote "single solute" and "interstitial", respectively. $E_i^{\text{Ni,int}}$ represents the minimized system energy when a Ni atom occupies an interstitial site *i*. $E_{\text{ref}}^{\text{Ni,bulk}}$ is the minimized system energy for a substitutional Ni atom in a chosen bulk site at the center of a large grain to avoid lang-range elastic interactions with GBs [54]. To balance the atom count, the average energy of Al in the bulk, $\bar{E}_{\text{ref}}^{\text{Al,bulk}}$, must be subtracted to account for the insertion of a Ni atom into an interstitial site at GBs:

$$\bar{E}_{\text{ref}}^{\text{Al,bulk}} = \frac{E^{\text{Al,bulk}}}{N}. \tag{2}$$

Here, $E^{\text{Al,bulk}}$ refers to the minimized total energy of the solvent Al atoms in bulk, and *N* is the total number of Al atoms in bulk. Then, molecular statics (MS) simulations were conducted using the CG algorithm. The per-site interstitial segregation energy was calculated using Eq. (1).

*2.3 Interstitial segregation energy with solute-solute interactions*

Solute segregation at GBs exhibits concentration dependence mainly due to the solute-solute interactions: repulsive interactions tend to suppress further segregation, while attractive interactions, conversely, enhance it [15,47,81]. Prior studies [64–67] have demonstrated that solute-solute interactions can significantly influence the accuracy of GB segregation predictions. Therefore, incorporating solute-solute interactions into the GB interstitial segregation energy may improve the reliability of GB segregation predictions.

In this study, we used the dual-solute (DS) segregation framework [66] to calculate the interstitial segregation energy accounting for solute-solute interactions. A neighbor list was constructed to include neighboring pairs of sites with a cutoff radius of 2.86 Å (approximately the nearest neighbor distance in Al): the potential interstitial site $i$ and its neighboring substitutional site $j$, where $j$ could be located either at a GB site or a near-GB bulk site. For simplicity, we assumed that solute-solute interactions between two interstitial sites are negligible. This assumption is based on the fact that sufficient hollow space is required for potential interstitial sites, making it unlikely for these interstitial sites to be neighbors. Further analysis and discussion of this simplification are provided in the Supplementary Materials. Therefore, the interstitial segregation energy with solute-solute interactions (denoted as $\Delta E_{i_j}^{\text{DS,int}}$) can be written as:

$$\Delta E_{i_j}^{\text{DS,int}} = E_{ij} - E_0 - \Delta E_j^{\text{Ni,sub}} - \bar{E}_{\text{ref}}^{\text{Al,bulk}} - 2\Delta E_{\text{ref}}. \tag{3}$$

where $E_{ij}$ is the minimized system energy with interstitial site $i$ and one of its neighbors substitutional site $j$ being occupied by two solute Ni atoms simultaneously. $E_0$ refers to the minimized system energy of a pure solvent system. $\Delta E_j^{\text{Ni,sub}}$ represents the SS segregation energy of substitutional site $j$, which is calculated by substituting site $j$ with a solute atom:

$$\Delta E_j^{\text{Ni,sub}} = E_j^{\text{Ni,sub}} - E_{\text{ref}}^{\text{Ni,bulk}}. \tag{4}$$

where $E_j^{\text{Ni,sub}}$ is the minimized system's energy with substitutional site $j$ occupied by a Ni atom. $\Delta E_{\text{ref}}$ in Eq. (3) is also a reference energy describing the difference between $E_{\text{ref}}^{\text{Ni,bulk}}$ and $E_0$ [66].

Finally, the DS interstitial segregation energy can be calculated through MS simulations using Eq. (3).

*2.4 Machine learning method*

Several machine learning (ML) models [55,56,67] have been developed to predict the substitutional segregation energies at GBs in NC metal alloys. These studies demonstrate the effectiveness of smooth overlap of atomic positions (SOAP) [82] descriptors in predicting substitutional segregation energies, suggesting that SOAP descriptors excel at capturing the essential features of local atomic environments (LAE) and provide valuable insights into the spectral nature of segregation energies. However, the ability of SOAP descriptors to predict interstitial segregation energies in substitutional binary alloys remains uncertain within the research community. To assess the predictability of interstitial segregation energy, we generated SOAP vectors using the DScribe Python package [83]. The parameters were set as $n_{max} = 12$, $l_{max} = 6$, and $r_{cut} = 6.0$ Å, which resulted in 2100 features for each interstitial site. Herein, we focus on the predictability of the SS interstitial segregation energy, which was randomly split into 50/50 test-train dataset. A linear regression algorithm [84] was employed to train the dataset and construct the ML model, chosen for its capability to deliver both rapid and precise training, along with reliable predictions of segregation energies [55].

**3 Hybrid MD/MC simulation results**

*3.1 Intraplanar interstitial segregation*

In this study, we define GB interstitial segregation as the phenomenon in which solute atoms segregate to interstitial sites at GBs without significantly altering their original structures. Based on this definition, Fig. 2 presents the GB structures and solute distributions for the Al-Ni system in several typical STGBs before and after hybrid MD/MC simulations at 300 K. Here, intraplanar interstitial segregation refers to the phenomenon in which interstitially segregated solute atoms remain within their corresponding solvent layers without causing significant distortion along the tilt axis directions. This is illustrated by the sliced side views in Fig. 2(c), (f), and (i).

For the Σ5(210) STGB model, the initial GB structure exhibits a periodic kite-like pattern, as shown in Fig. 2(a). This pattern can provide sufficient hollow space to accommodate some undersized metallic elements such as Cu, Ni, Co, and Fe [71]. After hybrid MD/MC simulations at 300 K with a solute concentration of 1 at.%, Ni atoms preferentially segregated to the interstitial sites within the hollow regions of the kite-like structure, rather than substituting the lattice sites of the kites. These segregated Ni atoms form a distinct single-layer segregation pattern, as indicated by the red arrow and dashed line in Fig. 2(b). Moreover, Fig. 2(b) and (c) demonstrate that the Ni atoms were aligned consecutively along $z$-direction and lying within the corresponding Al layers. This arrangement defines a specific segregation pattern, referred to as intraplanar interstitial segregation.

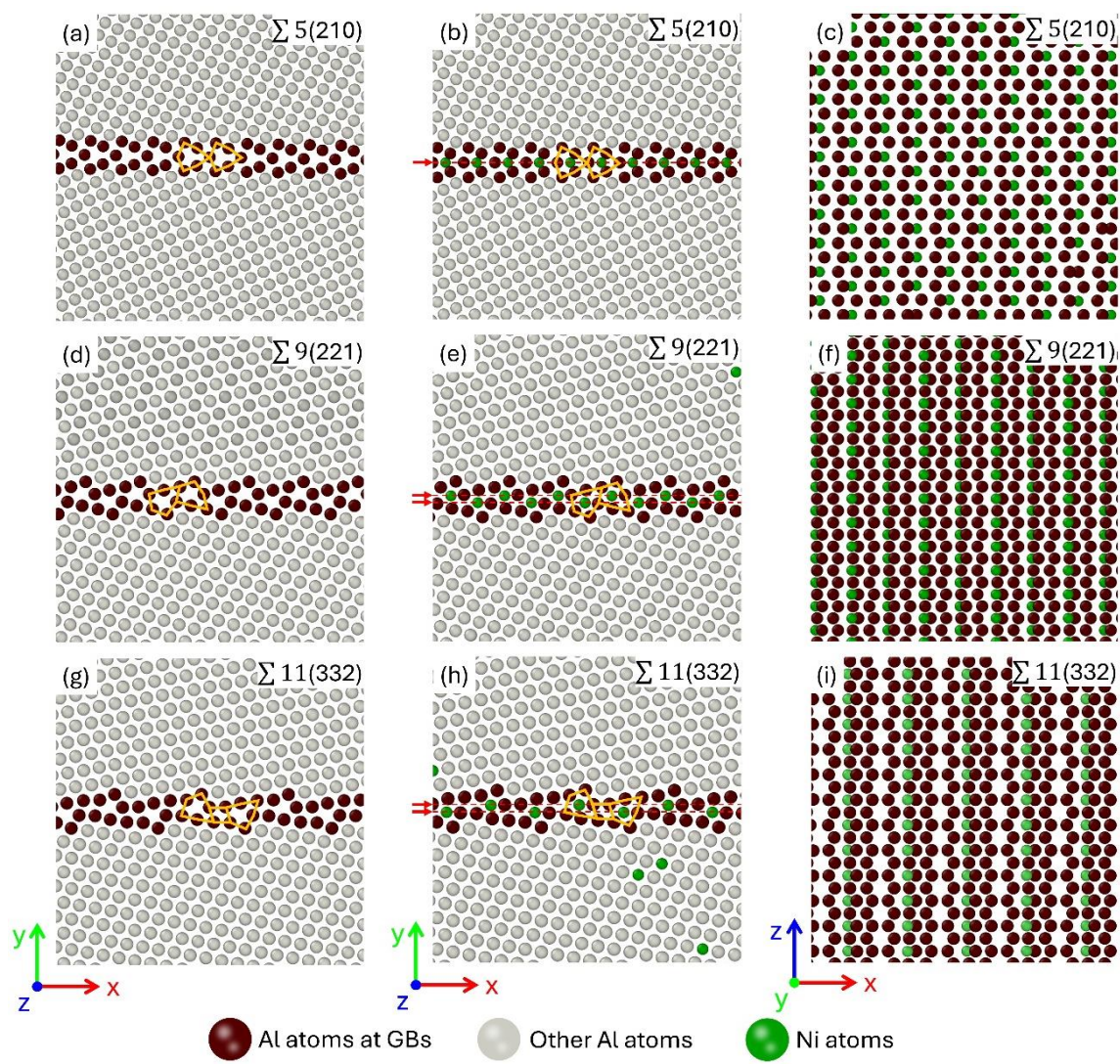

**Fig. 2** Hybrid MD/MC results at 300 K for the Al-Ni system of the Σ5(210), Σ9(221), and Σ11(332) STGBs with target solute concentrations of 1, 3, and 3 at.%, respectively. Panels (a), (d), and (g) show the top views (along the $z$-direction) of the Σ5(210), Σ9(221), and Σ11(332) STGBs prior to segregation. Panels (b), (e), and (h) display the corresponding top views after segregation. Meanwhile, panels (c), (f), and (i) present the side

views (along the *y*-direction) of the Σ5(210), Σ9(221), and Σ11(332) STGBs after segregation, with most bulk atoms removed for clarity.

Kite-like structures were also observed in the Σ9(221) and Σ11(332) STGBs, as shown in Fig. 2(d) and (g), respectively. After hybrid MD/MC simulations at 300 K with solute concentrations of 3 at.% in both cases, Ni atoms at the GBs preferentially occupied the interstitial sites within the free spaces of these kite-like structures. Notably, a bi-layer segregation pattern of Ni atoms formed in both Al STGBs, as indicated by the red arrows and dashed lines in Fig. 2(e) and (h). This bi-layer interstitial segregation pattern can be attributed to the extended periodicity of kite-like structures in these GBs.

Additionally, in the top views of the Σ5(210) and Σ9(221) STGBs, two adjacent interstitial Ni atoms were situated in two different Al layers, as shown in Fig. 2(c) and (f). This suggests that each Al layer likely contains an equal number of interstitial Ni atoms, and the periodicity of the Σ5(210) STGB corresponds to two kite-like structures. In contrast, for the Σ11(332) STGB, adjacent interstitial Ni atoms in the top view resided within the same Al layer, but the interstitial Ni layers were separated by an Al layer without any Ni atoms, as shown in Fig. 2(i).

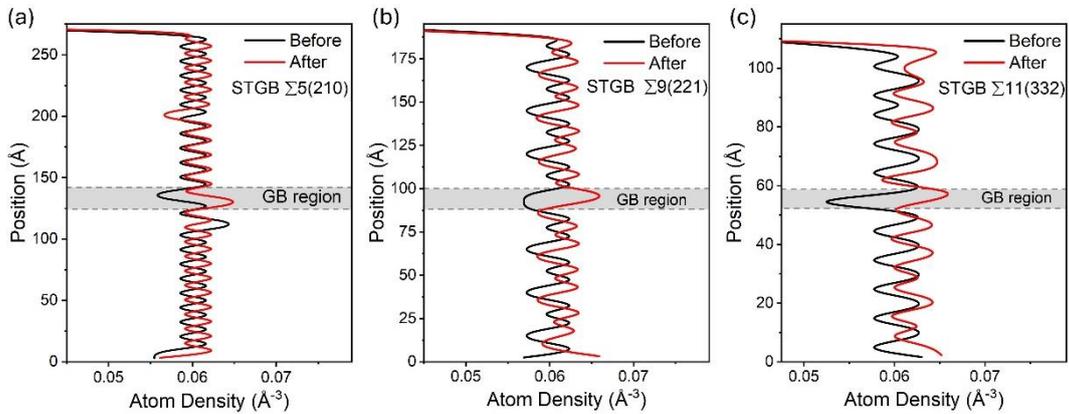

**Fig. 3** Comparison of atom density before and after hybrid MD/MC simulations at 300 K for the (a) Σ5(210), (b) Σ9(221), and (c) Σ11(332) STGBs, respectively.

Despite these differences, both scenarios observed in the Σ9(221) and Σ11(332) STGBs are classified as intraplanar interstitial segregation, as the segregated Ni atoms are confined to interstitial sites within the Al layers. Generally, these findings for the three representative STGBs are consistent with previous first-principles studies [68–71]. To verify that GB interstitial segregation occurred in the three typical STGBs, we calculated the atomic density along the *y*-direction, defined as the ratio of the number of atoms to volume, as shown in Fig. 3. The results reveal that, prior to segregation, the atom density at GBs is generally lower than the bulk density in all three cases. However, after segregation, the atomic density at GBs increases significantly, surpassing that of bulk regions. This confirms that GB interstitial segregation indeed occurred in these three representative STGBs after hybrid MD/MC simulations at 300 K. In addition to these findings, intraplanar interstitial GB segregation was also observed in certain Σ5, Σ9, and Σ11 asymmetric tilt GBs (ATGBs), as well as in some of the (001) and (110) STGBs, as illustrated in Supplementary Materials Fig. S2-7. It is important to note that the hybrid MD/MC simulations used in this study involved no atom insertion or deletion. Therefore, the observed interstitial segregation phenomena did not arise directly from atom insertion but instead resulted from GB evolution during the relaxation at finite temperatures.

*3.2 Interplanar interstitial segregation*

In contrast to intraplanar interstitial segregation, interplanar interstitial segregation describes the tendency of solute atoms to occupy the interstitial sites between solvent layers with obvious distortion along the tilt axis directions. This phenomenon was previously reported by Devulapalli et al. [33], who observed cage-like GB phase resulting from interplanar interstitial segregation of Fe in the Ti-Fe alloy system. In this work, we observed a similar interplanar interstitial segregation of Ni in Al, occurring alongside intraplanar interstitial segregation in specific CSL GBs. This behavior is exemplified in Fig. 4, which demonstrates the hybrid MD/MC results of the Al Σ5(9 13 0)/(310) ATGB at 300 K, with a solute concentration of 0.5 at.%. The original Al Σ5(9 13 0)/(310) ATGB exhibits a unique periodic facet kite-like structure, as illustrated in Fig. 4(a). Each periodic unit consists of seven kites: the sequence begins with a kite whose tip points to the upper left, followed by six kites pointing to the upper right.

After hybrid MD/MC simulations at 300 K with a solute concentration of 0.5 at.%, the overall GB structure and periodicity were largely preserved, as shown in Fig. 4(b). This top view reveals that most segregated Ni atoms preferentially occupied interstitial sites within the hollow spaces of the kite-like structures. This behavior is consistent with the segregation patterns observed in other ATGBs and STGBs of Al. The side view (Fig. 4(c)) further illustrates that most segregated Ni atoms were located within the Al layers, indicating the presence of intraplanar interstitial segregation. Interestingly, a notable phenomenon emerges: some segregated Ni atoms occupied the interstitial sites between Al layers. These interplanar segregated Ni atoms were highlighted using bright yellow arrows in Fig. 4(b) and (c). Unlike previous observations [33], these interplanar segregated Ni atoms were randomly distributed and appeared independent of one another. They were primarily found at or near the short facet of the periodic units, where the original kite-like structures transition into a circular configuration, as highlighted in Fig. 4(b). This phenomenon of interplanar interstitial segregation of Ni in Al is not confined to the current solute concentration, i.e., 0.5 at.%. It can also occur at higher concentrations, such as 3 at.%. Additionally, even under dilute conditions (e.g., 0.3 at.%), where Ni atoms do not fully occupy all intraplanar interstitial candidate sites at GBs, interplanar interstitial segregation may still take place, as demonstrated in Supplementary Materials Fig. S8. These findings suggest that the formation of interplanar interstitial segregation is independent of solute concentration and instead is governed by the intrinsic structure of GBs.

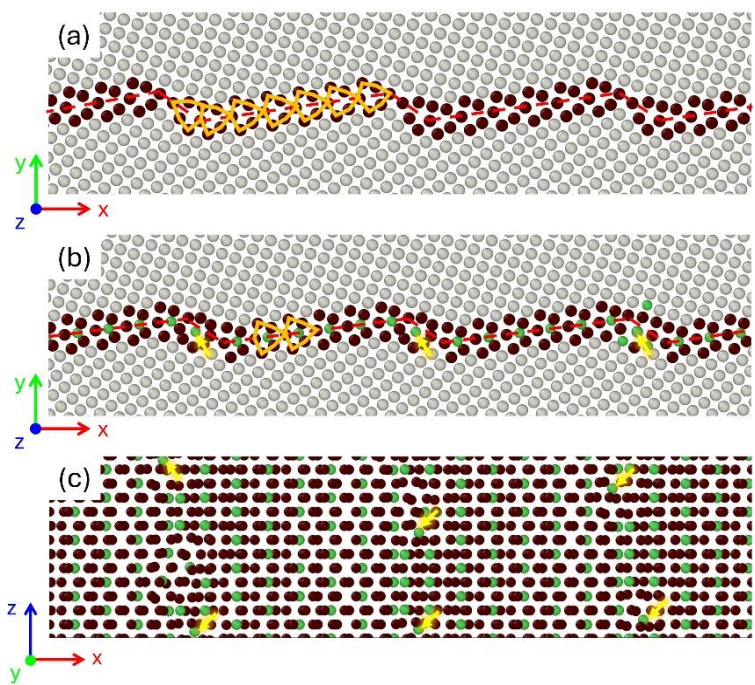

**Fig. 4** (a) Top view of the GB structure of the pure Al Σ5(9 13 0)/(310) ATGB. (b) Sliced view of the bottom layers of atoms in the z-direction after hybrid MD/MC simulations at 300 K with a solute concentration of 0.5 at.%. (c) Side view in the y-direction with most bulk atoms removed for better visualization. The faceting periodicity of the ATGB is emphasized with red dashed lines. Bright yellow arrows in (b) and (c) highlight the interplanar interstitial Ni atoms.

Based on these observations, a possible explanation for the formation of interplanar interstitial segregation within GBs is that Ni atoms segregate to an interstitial site whose free space is small than that of a perfect FCC Ni atom. These constrained interstitial sites are unable to fully accommodate the Ni atom, causing the surrounding GB atoms to displace the Ni atom into an interplanar interstitial site. In other words, interplanar interstitial segregation is caused by local GB structure distortion induced by solute segregation. Additionally, interstitial sites with larger free spaces are more favorable for Ni atom accommodation compared to smaller ones. This explains the low density of the interplanar interstitial segregation observed in the Al-Ni system across a broad range of concentrations. In addition to the Σ5(9 13 0)/(310) ATGB, interplanar interstitial segregation also occurred in other Al Σ5 ATGBs where local structures were distorted by solute atoms, as illustrated in Supplementary Materials Fig. S2-3, while no obvious interplanar interstitial segregation was observed in other CSL GBs. The overall density of interplanar interstitial solute atoms is relatively low compared to their intraplanar counterparts. Thus, while intraplanar interstitial segregation emerges as a predominant phenomenon in the Al-Ni system, it can occasionally be accompanied by interplanar interstitial segregation of solute atoms.

## 4 Interstitial site identification

### *4.1 Interstitial site identification in bicrystals*

Building on the observations of interstitial GB segregation in the Al-Ni system, a key question emerges: how can we effectively identify these interstitial sites? To address this, we began by analyzing the interstitial sites within the GBs of bicrystals. As previously noted, Ni atoms tended to segregate into the hollow regions of kite-like structures at GBs. Furthermore, previous studies [62,63,79] highlighted a strong correlation between segregation energy and free volume. Thus, having sufficient free space at a potential interstitial site is essential for accommodating a solute atom interstitially.

In the bulk regions of FCC metals, two types of well-defined interstitial sites are commonly recognized: octahedral sites (OS) and tetrahedral sites (TS). These sites can be obtained from the vertices of the Voronoi cell of each host atom. However, determining such interstitial sites becomes challenging in disordered regions, such as GBs. To solve this issue, Wagih and Schuh [85] proposed a Voronoi-based method to identify interstitial sites for hydrogen in NC Pd. Using the open-source software package Voro++ [86], they constructed Voronoi cells of the host atoms and considered their vertices as the potential interstitial sites. At GBs, vertices within a threshold distance of 1.0 Å were merged into distinct points, which were treated as the interstitial sites in these regions. To calculate the hydrogen trapping energy in FCC metals, Zhou et al. [87] used space tessellation method to determine the interstitial sites within a Σ5(130)[001] GB. Ding et al. [88] used a similar method to identify and calculate the favorable interstitial sites and the corresponding hydrogen trapping energy. Similarly, Ito et al. [89] introduced another Voronoi-based method to identify the interstitial sites in a body-centered cubic (BCC) Fe NC model. While these methods are effective for small elements, such as H, O, and N, they are not suitable for larger species, such as Ni, Cu, and Co. Therefore, a new approach must be developed to accurately identify interstitial sites for larger atoms.

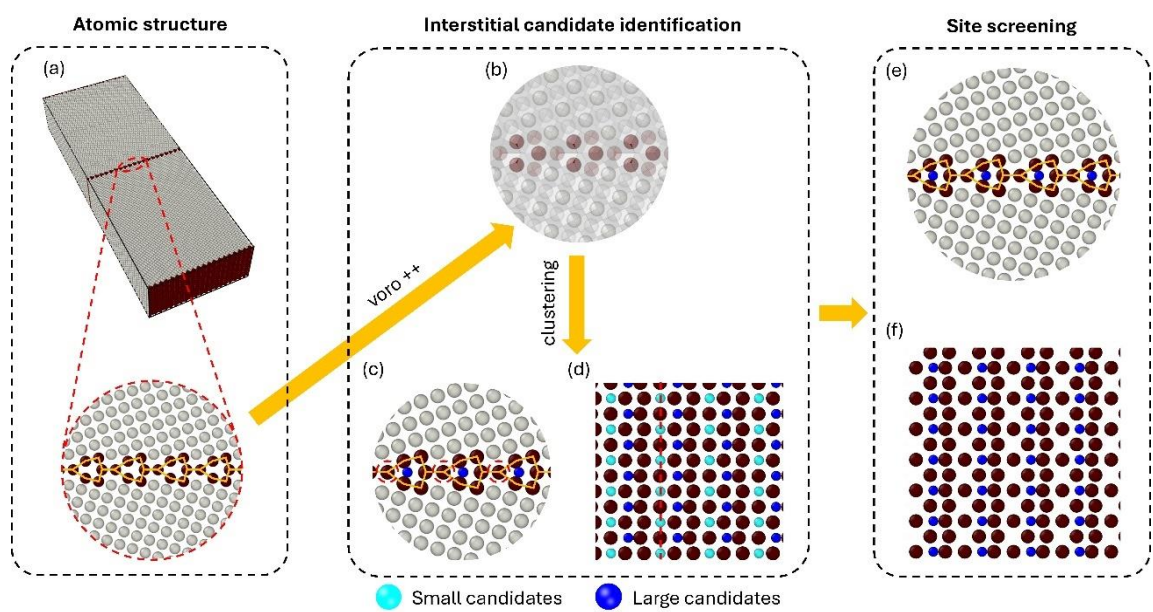

**Fig. 5** A schematic illustration of the flow path for interstitial site identification is shown as follows: (a) preparation of atomic structures including identification and separation of bulk and GB atoms; (b) generation of polyhedra for each atom, including both bulk and GB atoms; (c) and (d) illustrate the results of merging sites, where blue and cyan spheres represent interstitial site candidates; (e) and (f) show the site screening results. The grey spheres represent FCC atoms, while the dark red spheres indicate GB atoms. The circular views provide the top-down perspectives of GBs, whereas the rectangular views offer side perspectives with FCC atoms removed for better visualization.

In this study, we enhanced the Voronoi-based approach proposed by Wagih and Schuh [85] to identify interstitial sites for metallic elements. The workflow for this process is illustrated in Fig. 5. The method began with structure identification and separation of the optimized bicrystal using the a-CNA algorithm [76], which is essential for subsequent interstitial site screening. As shown in Fig. 5(a), the initial pristine Al bicrystal GB features periodic kite-like structures. Voronoi polyhedra were then generated for each atom in the system using the Voro++ package [86]. In a perfect FCC unit cell, nearest neighbors share mutual vertices due to the long-range order of atomic arrangements, which allows these vertices to be directly considered as interstitial sites for small elements, such as H, O, and N. However, at GBs, the atomic arrangements become disordered, which induces the vertices of neighboring atoms being separated yet remain in close to one another.

To account for this, a DBSCAN clustering algorithm [84] was employed to merge these nearby vertices into single point. This clustering algorithm automatically determines the number of clusters based on only two key parameters and is capable of effectively handling large size 3D spatial data. The cutoff distance ($d_c$) and minimum sample size ($N_{min}$) are two key parameters that significantly affect the interstitial site identification results. $d_c$ defines the maximum distance between two points for them to be considered part of the same cluster. Small $d_c$ will generate more interstitial candidates, while large value may result in fewer candidates. For the Al-Ni system, $d_c$ typically ranges from 1.0 to 1.6 Å. Setting $d_c$ below 1.0 Å leads to an excessive number of calculations, whereas values above 1.6 Å may cause failures in candidate site identification. In this study, we chose $d_c = 1.25$ Å, which is approximately the radius of a Ni atom. $N_{min}$ determines the minimum number of points (including the point itself) that must be within the $d_c$-radius neighborhood for a point to be considered a core point. Suitable values for $N_{min}$ typically range from 1 to 12, with $N_{min} = $ 5, 6, or 7 being the most effective choices in Al-Ni bicrystals based on our tests. The clustering results are shown in Figs. 5(c) and (d), where two types of interstitial candidates are marked by cyan and blue spheres, respectively. The cyan spheres are located below the kite-tips in the top view, as highlighted by red dashed circles in Fig. 5(c), while the blue ones are positioned at the core of the kites. It was observed that the cyan spheres are concentric with the kite-tip Al atoms, as illustrated by the red dashed line in Fig. 5(d). In contrast, the kite-core candidates remain uncovered by Al atoms. This observation led to the hypothesis that the small candidates would have smaller free volumes compared to the blue ones. This hypothesis was validated through Voronoi volume calculations, which revealed that the free volume of the small candidates (~9.61 Å$^3$) is indeed smaller than that of the large candidates (~11.96 Å$^3$). These smaller sites are more likely to induce significant distortion and alter the GB structure upon Ni segregation. To minimize such effects, candidates with free volumes ($V_f$) smaller than ~10.8 Å$^3$ ==(approximately the atomic volume of perfect FCC Ni, calculated using its lattice constant of 3.52 Å.)== should be removed. The site screening results, shown in Figs. 5(e) and (f), confirm that the small candidates have been eliminated, leaving only the larger candidates at the centers of the kite-like structures.

*4.2 Validation in bicrystals*

Using the enhanced Voronoi-based method, we sought to identify the potential interstitial sites in several bicrystals. The results for the three abovementioned STGBs are shown in Supplementary Materials Fig. S9. These findings indicate that the distribution of identified interstitial sites matches that of the hybrid MD/MC results displayed in Fig. 3. Moreover, it is worth noting that the interstitial sites identified before and after the site screening process are identical. This indicates that only one type of interstitial site has been identified during the initial site identification process, which is probably due to the high symmetry of the GB structures, making the site screening process unnecessary for these STGBs. However, this does not imply that site screening is unnecessary for all STGBs. For instance, in the case of Σ5(310), two types of interstitial sites are initially identified, and the screening process successfully eliminates the small interstitial sites, as shown in Fig. 5.

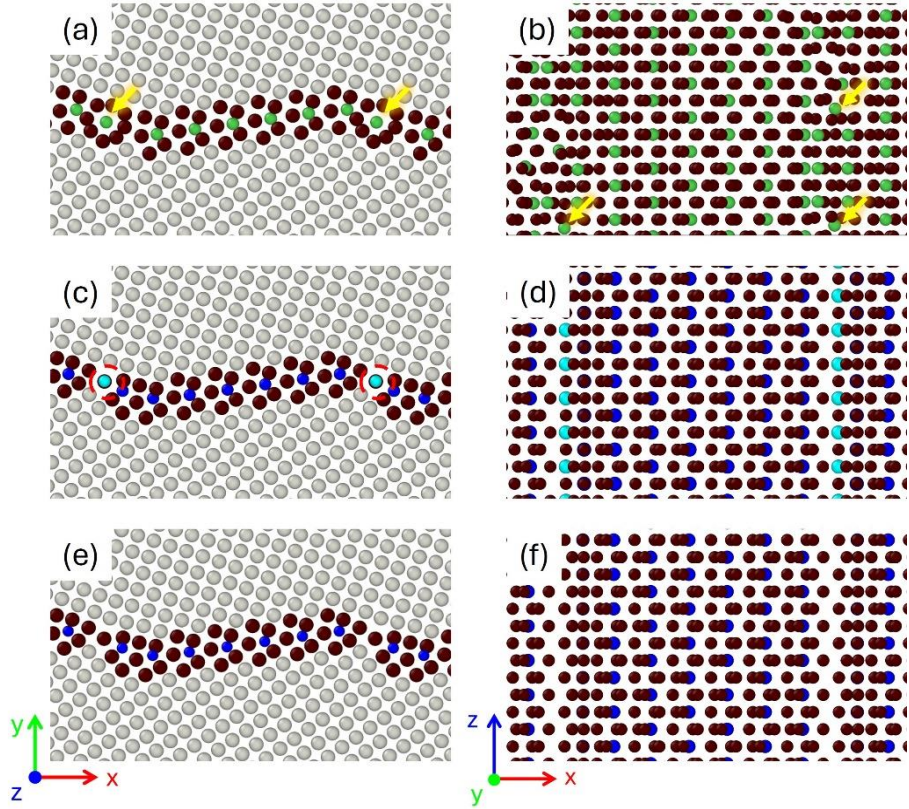

**Fig. 6** Comparison between identified sites and hybrid MD/MC simulations for the Al Σ5(9 13 0)/(310) faceting ATGB. (a) and (b) show the distribution of segregated Ni atoms after hybrid MD/MC simulations at 300 K with a solute concentration of 0.5 at.%. The bright yellow arrows represent interplanar interstitials. (c) and (d) are the results of initial interstitial site identification, where blue spheres are large sites, while the cyan ones are small sites. (e) and (f) are results of site screening. The left panels are top views showing a face unit, while the right panels are side views in *y*-direction with bulk atoms removed to enhance clarity.

Fig. 6 compares the distribution of interstitial Ni atoms obtained from hybrid MD/MC simulations at 300 K with those identified using the enhanced Voronoi-based method. Most segregated Ni atoms occupied interstitial sites within the kite-like structures, exhibiting intraplanar interstitial segregation, as shown in Fig. 6(a). Only a few Ni atoms were located at those between-plane interstitial sites, corresponding to interplanar interstitial segregation, as illustrated in Fig. 6(b). After the initial site identification process, two types of interstitial sites were determined, as shown in Fig. 6(c) and (d). The cyan spheres correspond to smaller interstitial sites with a free volume of ~10.06 Å³, which is less than the free volume of perfect FCC Ni, while the blue spheres represent large interstitial candidate sites with free volumes of ~12.27 Å³ or more. Following the site screening process, as shown in Fig. 6(e) and (f), the smaller interstitial sites have been eliminated, leaving only the large interstitial candidate sites.

Interestingly, Fig. 6(a) reveals that interplanar interstitial segregation occurs at the short facets of the periodic GB units, the same locations where smaller interstitial sites were identified in Fig. 6(c). This supports the hypothesis that Ni atom absorption at smaller interstitial sites can lead to interplanar segregation due to significant GB structure distortion caused by segregation. In addition to this, the distribution of interstitial sites after screening closely aligns with the intraplanar segregation pattern of Ni atoms shown in Fig. 6(a) and (b). These results demonstrate that the enhanced Voronoi-based method can effectively identify the interstitial candidate sites within these CSL GBs.

**5 GB interstitial segregation in NC metals**

*5.1 Interstitial sites in NC*

Large free spaces may also be present in NC metals due to the disorder within the GB network, particularly in junction regions. Consequently, GB interstitial segregation can occur in NC metals. To identify interstitial sites within the GB network, we utilized the fully relaxed Al-16 model. The initial identification of interstitial sites is depicted in Fig. 7(a) and (b). A total of 11,583 interstitial sites were identified following the vertices clustering step shown in Fig. 5. The atomic volumes of these sites range from approximately 5.5 to 15.5 Å$^3$, as shown in Fig. 7(a). As previously discussed in bicrystals, many of these small interstitial sites are unsuitable for absorbing Ni atoms due to the resulting significant local atomic strain, which would increase the system energy. Therefore, we adopted a volume screening criterion of 10.0 Å$^3$ for the Al NC model, slightly smaller than the value of 10.8 Å$^3$ used for bicrystals, to include as many interstitial candidate sites as possible while permitting minor distortions.

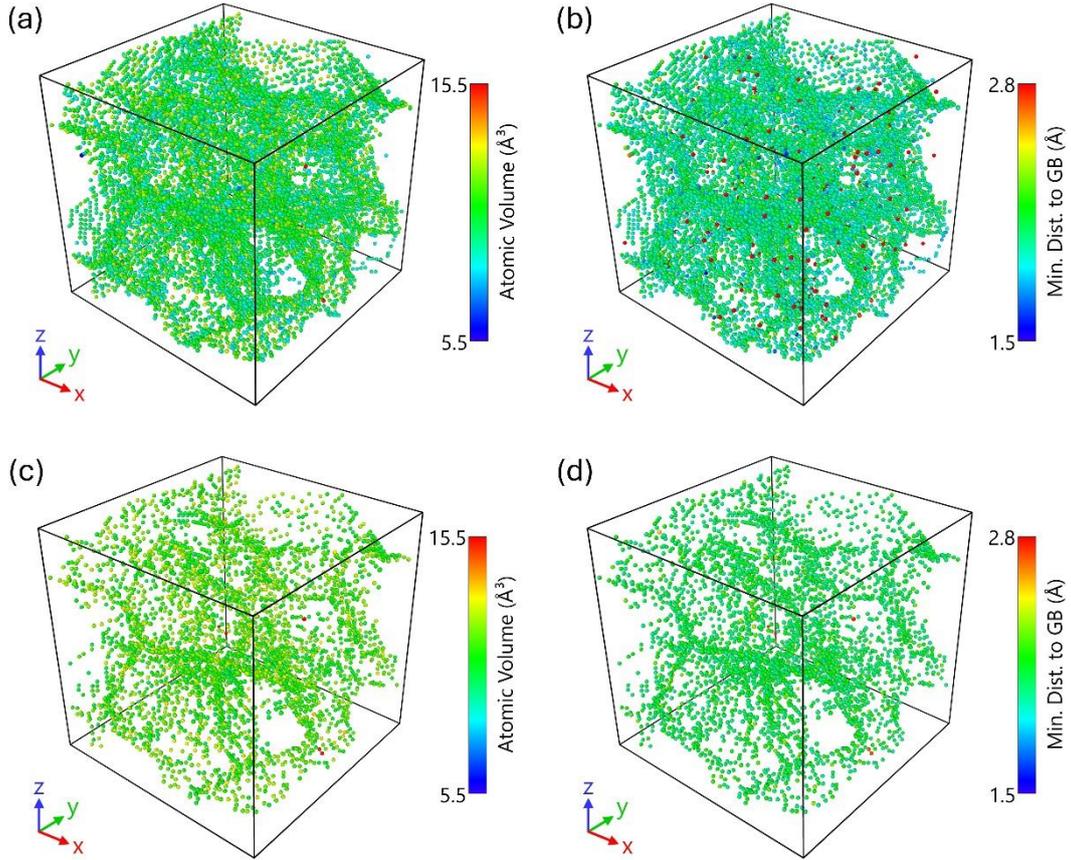

**Fig. 7** (a) and (b) show the atomic volume and minimum distance to GB atoms of initial identification results. (c) and (d) demonstrate the corresponding site screening results.

However, we found that applying a single volume screening criterion was insufficient. Figure 7(b) shows the minimum distance from interstitial sites to GB atoms, revealing that some of these sites overlap significantly with solvent atoms. Such overlap could lead to large atomic displacements during MS simulations, which may cause a considerable increase in system energy. To address this, a new screening criterion based on the minimum distance to GB atoms ($d_{min}$) was introduced, with the threshold set to 2.0 Å to minimize atomic displacements. Moreover, we excluded the nearest neighboring interstitial sites within 2.5 Å (approximately the diameter of a Ni atom) by assigning the large hollow space, if present, to a single interstitial site. After this additional filtering, 4,149 interstitial sites were identified. Figures 7(c) and (d) display the atomic volume distribution and minimum distance to GB atoms for the filtered sites. Notably, the final interstitial sites are randomly distributed across the GBs, without significant clustering observed in triple junctions due to the site screening criteria. These parameters can be adjusted as needed. Larger $V_f$ and $d_{min}$ will reduce the number of identified interstitial sites, which in turn decreases the dataset size for segregation energy calculations and may introduce slight deviations in segregation predictions.

Previous interstitial site identification methods [85,87,89] tend to generate an excessively high number of interstitial sites, as they were primarily designed for identifying interstitial sites for hydrogen in metals. These methods are not well-suited for substitutional alloy systems, where the larger atomic size of solute atoms makes sufficient free space within GBs a critical factor. In contrast, the enhanced Voronoi-based method developed in this study is primarily designed for identifying interstitial sites in substitutional alloy systems. It effectively detects interstitial sites in both bicrystals and NC systems. Additionally, with minor parameter adjustments, this method can also be applied to identify interstitial sites for smaller elements such as H, O, and N, which will be discussed later.

*5.2 Interstitial segregation energies*

After identifying the potential interstitial sites in the Al-16 model, we conducted MS simulations to calculate the SS and DS interstitial segregation energies using Eqs. (1) and (3), respectively. The SS interstitial segregation energy represents the interstitial segregation energy of a single solute without solute-solute interactions, while the DS interstitial segregation energy accounts for the interactions between an interstitial solute and its nearest neighboring substitutional solute. It is worth noting that interstitial atoms as nearest neighbors are excluded based on the applied interstitial site screening criteria. To evaluate the influence of interstitial segregation energies, we combined the SS interstitial segregation energy ($\Delta E_i^{\text{SS,int}}$) with the DS substitutional segregation energy $\Delta E_i^{\text{DS}}$ from our previous work [66], creating the DS-SS-int segregation energy dataset. This dataset includes SS and DS substitutional segregation energies as well as the SS interstitial segregation energy. Additionally, a full segregation energy dataset was obtained by incorporating the $\Delta E_i^{\text{SS,int}}$, $\Delta E_i^{\text{DS,int}}$, and $\Delta E_i^{\text{DS}}$ together, including all types of segregation energies, i.e., SS and DS substitutional segregation energies, and the SS and DS interstitial segregation energies.

Fig. 8(a) shows the spectral distributions of the DS-SS-int and full segregation energy datasets. Similar to the substitutional segregation energy spectra [55], both exhibit skew-normal distributions. When fitted using the skew-normal model [54], the spectra show very similar shapes and positions, as reflected in the fitting parameters $\alpha$, $\mu$, and $\sigma$, which describe the shape (i.e., skewness of the distribution), characteristic energy (i.e., position of the distribution), and width of the skew-normal distribution, respectively. The positive $\alpha$ values indicate a slight leftward skew in the spectra. These values are close to, but slightly smaller than, the $\alpha$ value of 1.52 observed for the DS spectrum [66]. This similarity suggests that neither SS nor DS interstitial segregation energies can significantly alter the skew direction of the spectrum. Instead, they only slightly reduce the skewness, indicating an increase in the population of positive interstitial segregation energies within the datasets. This trend in skewness is further reflected in the slight increase of $\mu$ values.

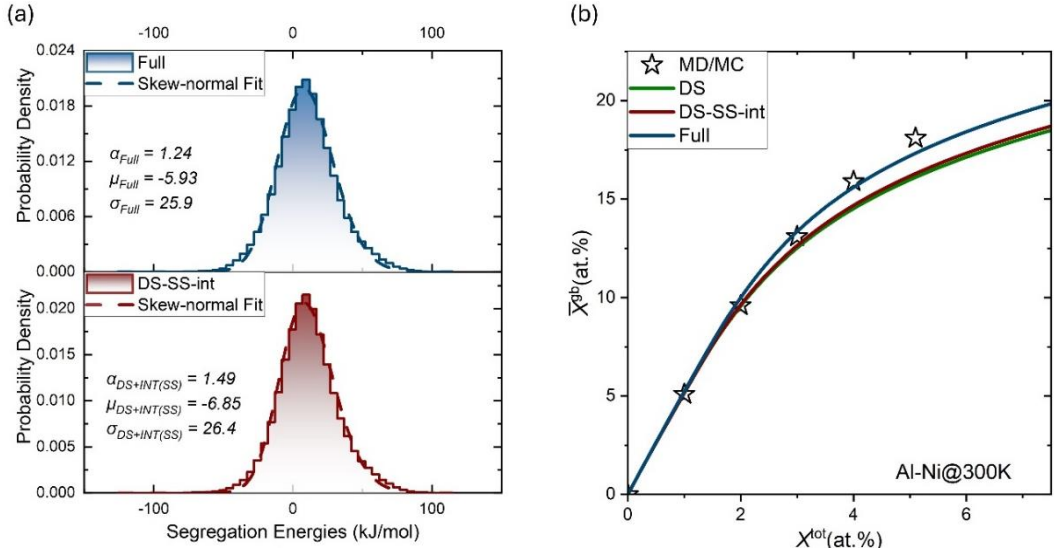

**Fig. 8** (a) presents the segregation energy spectra for DS-SS-int and full cases. (b) demonstrates the effect of interstitial segregation energies on segregation prediction accuracy. The hybrid MD/MC results at 300 K, along with the DS prediction curve (green), are adapted from Ref. [66].

To evaluate the effects of SS and DS interstitial segregation energies on segregation predictions, the DS model [66] was numerically solved for GB solute concentration ($\bar{X}^{gb}$) as a function of total solute concentration ($X^{tot}$), as shown in Fig. 8(b). Compared to DS substitutional segregation energy spectrum, the DS-SS-int spectrum slightly increased the prediction accuracy relative to the hybrid MD/MC results at 300 K. This improvement indicates that incorporating the SS interstitial segregation energy can positively influence the segregation predictions, despite only 4149 interstitial sites being identified. Additionally, the full spectrum demonstrates even better alignment with the hybrid MD/MC results, highlighting that interstitial segregation energy, when accounting for solute-solute interactions, can significantly enhance segregation prediction accuracy.

It is essential to illustrate the distribution of interstitial sites with negative interstitial segregation energies. To achieve this, we present a color-coded map of solute atoms at these sites in Supplementary Fig. S10. The map reveals that those favored interstitial sites are well distributed throughout the entire GB network. This distribution is as expected owing to the random orientation of grains in the nanocrystalline sample, which results in a high degree of GB randomness. These findings highlight the importance of understanding and predicting interstitial segregation in in substitutional alloys.

## 6 Discussions

### *6.1 Predictability*

In a substitutional alloy system, when a solute atom segregates to an interstitial site at GBs, the LAE of its surrounding atoms will be influenced by some intrinsic interactions, such as elastic and chemical effects. These interactions serve as distinctive fingerprints to assess the favorability of segregation at the corresponding interstitial site. However, upon relaxation, the solute atom may shift from its original position (i.e., the identified interstitial site). For example, Supplementary Materials Fig. S11 illustrates the distribution of atomic displacements after MS simulations under zero-pressure conditions for the Al-Ni system in the Al-16 NC model. The average displacement was approximately 0.139 Å, which corresponds to about 5.6% of the Ni atomic diameter. Such displacements may reduce the predictability of interstitial segregation energy when using SOAP descriptors.

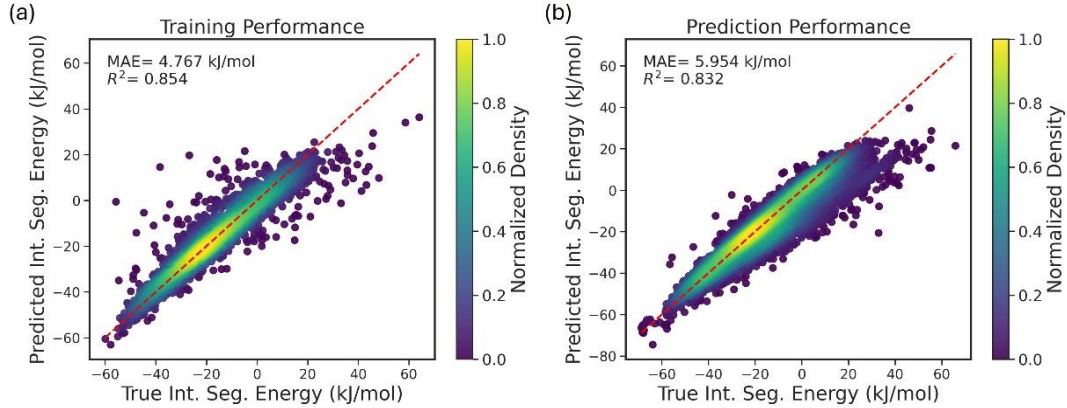

**Fig. 9** Color mapping of the true vs predicted interstitial segregation energies in the Al-Ni system: (a) training performance of the ML approach in the Al-16 NC model, and (b) prediction performance in a large NC Al model ($20^3$ $nm^3$) by the trained model.

For the Al-Ni system with the size of $16^3$ $nm^3$, the ML training performance is shown in Fig. 9(a). The results reveal that the predicted interstitial segregation energies closely match the true values, with an R-squared value of 0.854 and a mean absolute error (MAE) of 4.767 kJ/mol. These metrics are reasonable given the wide range of interstitial segregation energy, which spans from -60 to 60 kJ/mol. The color map of normalized density highlights the concentration of data points, with brighter regions signifying higher probability of interstitial segregation energy. Notably, Fig. 9(a) shows that most data points are clustered around -20 kJ/mol, indicating that most identified interstitial sites at GBs are favorable for accommodating Ni atoms, and confirming the reliability of the enhanced interstitial site identification method. These findings demonstrate that SOAP descriptors are powerful to predict the interstitial segregation energy in the Al-Ni system, even under significant atomic displacements observed during MS simulations.

We also performed interstitial site identification and MS simulations for a large NC Al model ($20^3$ $nm^3$), obtained from the substitutional segregation energy database developed by Wagih and Schuh [55]. The interstitial segregation energy of this larger model was used to evaluate the predictive performance of the trained ML model. Fig. 9(b) shows the true vs predicted interstitial segregation energy which was predicted using the SOAP descriptors of the large NC structure. The results show an R-squared value of 0.832 and a MAE of 5.954 kJ/mol. Similar to the Al-16 NC structure, the dense region of data points is located around -20 kJ/mol. These findings confirm the reliability of the trained

ML model and the robustness of SOAP descriptors in predicting interstitial segregation energy. Furthermore, this approach demonstrates the capability to predict interstitial segregation energy for larger NC models using data derived from smaller systems.

Additionally, we constructed two sizes ($16^3$ and $20^3$ $nm^3$) of NC Pd structures using the Atomsk software package [90]. The EAM potential developed by Zhou et al. [91] was employed to evaluate the interatomic interactions for the Pd-H system. Following full relaxation, we applied the interstitial site identification method, while adjusting parameters such as the minimum cluster size, the minimum distance to GB atoms, and the neighboring distance. Unlike previous studies [85,89,92], our analysis focused on the interstitial segregation energies of hydrogen in Pd at those interstitial sites specifically located within GB cores, while excluding near-GB interstitial sites. This approach assumes that GB cores offer more free space compared to regions adjacent to the GBs. The training and prediction performances of the ML model using SOAP descriptors are presented in Supplementary Materials Fig. S12. The results further demonstrate the applicability of the interstitial site identification method to small interstitial species and demonstrate the effectiveness of SOAP descriptors in predicting interstitial segregation energies in such systems. Nevertheless, we need to emphasize that these ML models are specifically focused on predicting the SS interstitial segregation energy. Further research is required to develop effective ML models for accurately predicting DS interstitial segregation energies in NC alloys.

*6.2 Interstitial segregation-induced GB transition*

Previous observations indicate that a kite-like structure is a necessary condition for GB interstitial segregation in substitutional alloys. However, the presence of a kite-like structure does not necessarily guarantee that GB interstitial segregation will occur. For instance, in the Al Σ13(510) STGB, potential interstitial sites with free volumes of approximately 9.996 Å³ (small) and 11.917 Å³ (large) can initially be identified within the cores of these kite-like structures. After screening, the large interstitial candidates remain. Nevertheless, hybrid molecular MD/MC simulations reveal that Ni atoms preferentially segregate to substitutional sites rather than interstitial ones, as illustrated in Supplementary Materials Fig. S13. This observation aligns with previous first-principles findings [71], which attribute the behavior to the loose-packed nature of the GB structure and its relatively high GB energy.

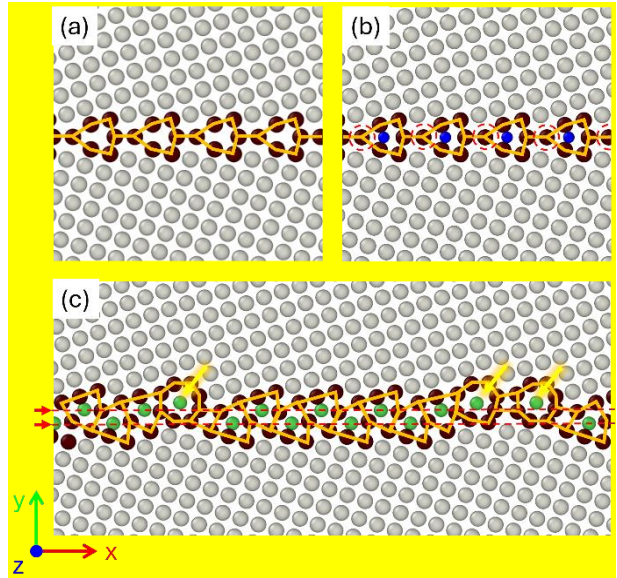

**Fig. 10** (a) A pure Al Σ5(310) STGB prior to segregation for reference. (b) Identified interstitial sites (blue spheres) determined using the enhanced Voronoi-based method. (c) A sliced view in the *z*-direction showing two layers of atoms after hybrid MD/MC simulations with a solute concentration of 0.5 at.% at 300 K. The red arrows and red dashed lines illustrate the bi-layer interstitial segregation pattern.

Interstitial sites were also identified within the Σ5(310) STGB, as shown in Fig. 10(b). After hybrid MD/MC simulations at 300 K, interstitial segregation occurred, unexpectedly resulting in a kite transition, as shown in Fig. 10(c), rather than the anticipated single-layer interstitial segregation displayed in Fig. 10(b). Most Ni atoms segregated to interstitial sites and formed a two-layer interstitial segregation pattern, indicating the transition of kite structures. Additionally, randomly distributed interplanar interstitial segregation of Ni (highlighted by yellow arrows) was observed, indicating the coexistence of intraplanar and interplanar interstitial segregation behaviors in the system. These interplanar interstitial segregation units have the same structure as those within the Σ5(9 13 0)/(310) ATGB shown in Fig. 4. We attribute these interstitial segregation-induced GB transition and the interplanar interstitial segregation transition to the existence of small interstitial sites between Al atoms in *z*-direction, highlighted using the red dashed circle in Fig. 10(b) and Fig. 6(c) and (d). These observations differ from previous first-principles studies [68,70,71], who reported that the interstitial sites among kite cores were the most favorable sites for undersized species, such as Ni and Cu. This discrepancy is likely due to the zero-temperature conditions in first-principles studies, which do not account for significant GB evolution.

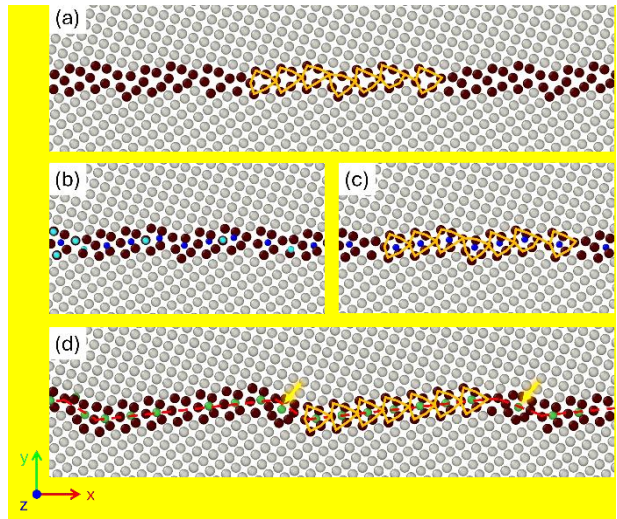

**Fig. 11** Illustration of interstitial segregation-induced faceting in the Al Σ5(230)/(17 6 0) ATGB. (a) Initial Gb structure before segregation. (b) Sliced view of identified interstitial candidate sites in the initial GB structure. Cyan spheres are small candidates, while blue spheres are large candidates. (c) Identified interstitial sites after site screening. (d) Interstitial segregation-induced faceting after hybrid MD/MC simulations at 300 K with a solute concentration of 0.5 at.%. The bright yellow arrows indicate the interplanar interstitial Ni atoms.

In the Al Σ5(230)/(17 6 0) ATGB, we observed a remarkable transition: interstitial segregation-induced GB nano-faceting after hybrid MD/MC simulations at 300 K, as shown in Fig. 11. The original Σ5(230)/(17 6 0) ATGB contains only short facets, as illustrated in Fig. 11(a), which contrasts with the Σ5(9 13 0)/(310) faceting ATGB. During the initial site identification process, small interstitial candidate sites were identified, highlighted as cyan spheres in Fig. 11(b). These sites were subsequently removed during the filtering process, as shown in Fig. 11(c). After interstitial segregation, long-range facets formed, accompanied by interplanar interstitial segregation, as indicated by the bright yellow arrows in Fig. 11(d). This observed interstitial segregation-induced faceting is different from previously reported substitutional segregation-induced nano-faceting in the Cu-Ag system [31], where the solvent atoms were significantly smaller than the solute atoms.

Generally, these interstitial segregation-induced GB transitions are probably caused by the presence of small interstitial sites, which allow to adsorb Ni atoms but causing significant local atomic distortion or rearrangements. Unlike previously observed segregation-induced GB transitions at high temperatures [29,31], the GB transitions in this study occurred at room temperature. No split-kite transition was observed after hybrid MD/MC simulations at 300 K. Overall, while the enhanced Voronoi-based interstitial site identification method may not fully capture these GB transitions or non-interstitial segregation behaviors within kite-like structures, the detection of small interstitial sites could serve as an indicator of potential GB transitions.

Hybrid MD/MC simulations revealed that intraplanar interstitial segregation was more commonly observed than interplanar interstitial segregation in Al-Ni bicrystals. However, interplanar interstitial segregation remains a critical phenomenon, as it often occurs alongside GB phase transitions, especially in specific Σ5 GBs. Additionally, interstitial site identification results suggest that interplanar segregation tends to form in regions where the available free space is slightly smaller than that of intraplanar sites. In this context, structural relaxation may induce significant local distortions and even trigger structural transitions. For NC systems, incorporating slightly smaller interstitial sites may enhance the accuracy of segregation predictions. This explains why we used $V_f = 10.0 \text{ Å}^3$ instead of 10.08 $\text{Å}^3$ for interstitial site screening in the NC Al samples.

*6.3 Temperature independence and interstitial segregation in other systems*

To assess the effect of temperature on interstitial segregation, we performed hybrid MD/MC simulations at 400 K and 500 K in the Al STGB Σ5(210) bicrystal employing the EAM potential [77]. A constant chemical potential of $\Delta\mu_0 = 2.165$ and a target solute concentration of Al-Ni 1 at.% were used to sample the alloy system. The results showed clear interstitial segregation at both temperatures, as illustrated in Supplementary Fig. S14, indicating that temperature does not significantly affect the interstitial segregation behavior of Ni in Al bicrystals up to moderate temperature conditions.

To eliminate the influence of interatomic potentials, we employed two additional EAM potentials [93,94] for the Al-Ni system and performed hybrid MD/MC simulations at 300 K in the STGB Σ5(210) bicrystal. The results confirmed that interstitial segregation occurred in both cases, as illustrated in Supplementary Fig. S15. Furthermore, interatomic potentials for the Ag-Cu [95] and Ta-Cu [96] systems were used to investigate Cu segregation behavior in FCC Ag-Cu and BCC Ta-Cu alloy systems.

The results showed that interstitial segregation also occurred after hybrid MD/MC simulations at 300 K, as presented in Supplementary Fig. S16. Notably, these interstitially segregated solute species are all smaller in atomic size compared to their solvent species. In contrast, solutes with equivalent or larger atomic sizes, such as Mg in an Al matrix, only occupy substitutional sites in GBs at room temperature, as shown in Supplementary Fig. S17. These findings align with previous first-principles studies [68–71], further reinforcing the observations in this study.

## 7 Conclusion

In summary, we observed GB intraplanar and interplanar interstitial segregation phenomena in the Al-Ni substitutional alloy system through hybrid MD/MC simulations at finite temperatures. Guided by the observations and analyses from atomistic simulations, we developed an enhanced Voronoi-based method to identify interstitial candidate sites in both CSL GBs and NC regular GBs. This method enabled the calculation of SS and DS interstitial segregation energies. Additionally, linear ML models were developed to assess the predictability of SS interstitial segregation using SOAP descriptors. The key conclusions are drawn as follows:

- GB intraplanar interstitial segregation was prominently observed within numerous CSL GBs featuring kite-like structures after hybrid MD/MC simulations at 300 K in the Al-Ni system. In contrast, interplanar interstitial segregation behaviors were primarily detected in Σ5 asymmetric tilt ATGBs due to their intrinsic GB structures, although their density was typically lower compared to intraplanar segregation.
- Interstitial segregation can induce unexpected GB transitions, such as kite transitions and nano-faceting. Unlike GB structural transitions typically reported in substitutional segregation at elevated temperatures, the transitions observed from this work were closely linked to the presence of kite-like structures and small interstitial candidate sites which can be effectively identified using the enhanced Voronoi-based method.
- The enhanced Voronoi-based method effectively identified interstitial sites in both NC substitutional and interstitial alloy systems. While the computed SS interstitial segregation energy slightly improved segregation prediction accuracy, the DS interstitial segregation energy significantly enhanced the accuracy of predicting segregation behaviors in the Al-Ni system.

- The ML results demonstrated that SOAP descriptors effectively captured the LAE of interstitial sites in both Al-Ni and Pd-H systems. Linear ML models trained on small structures accurately predicted per-site interstitial segregation energies in larger alloy systems, enhancing the scalability and applicability of the approach.

However, due to the inherent randomness and significant evolution of regular GBs at finite temperatures, distinguishing between substitutional and interstitial segregated solute atoms in NC substitutional alloys remains challenging and requires further investigations. Overall, our study highlights the critical role of GB interstitial segregation in substitutional binary alloys and provides valuable insights for advanced alloy design.

**Acknowledgement**

This research was supported by the NSERC Discovery Grant (RGPIN-2019-05834), Canada, and the use of computing resources provided by the Digital Research Alliance of Canada. Z.Z. acknowledges financial support from the University of Manitoba Graduate Fellowship (UMGF). During the preparation of this manuscript, the authors used ChatGPT to improve its readability. The authors carefully reviewed and edited the manuscript following the use of this tool and took full responsibility for the content of the publication.

**Declaration of Competing Interest**

The authors declare that they have no known competing financial interests or personal relationships that could influence the work reported in this paper.

**Code availability**

The source code of the interstitial site identification method introduced in this paper is now available in the GitHub repository (https://github.com/ZacharyZhang99/interstitial_site_identification).

# Supplementary materials for

# Grain boundary interstitial segregation in substitutional binary alloys

Zuoyong Zhang and Chuang Deng*

*Department of Mechanical Engineering, the University of Manitoba, Winnipeg, Canada MB R3T 5V6*

* Corresponding author: Chuang.Deng@umanitoba.ca

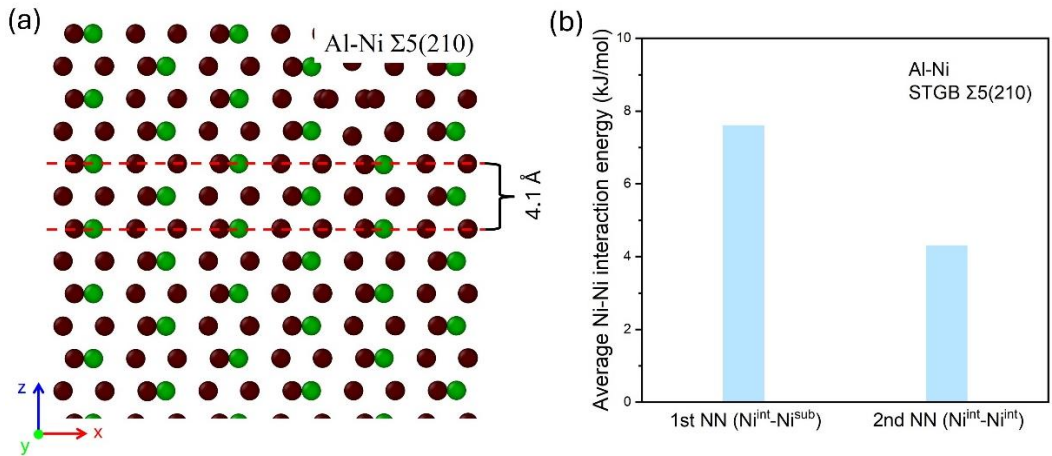

**Fig. S1** (a) A sliced side view illustrating the ~4.1 Å distance between the nearest segregated interstitial Ni atoms along the z-direction. (b) A comparison of the average Ni-Ni interaction energies between an interstitial Ni and its substitute first nearest neighbors (1st NNs) versus its nearest interstitials (2nd NNs).

In this study, we focus exclusively on solute-solute interactions involving interstitial sites and their nearest neighboring substitutional atoms. Interstitial Ni atoms may experience elastic interactions along the *z*-direction (i.e., the tilt axis). In the segregated $\Sigma 5(210)$ Al bicrystal, the distance between two adjacent interstitial Ni atoms is approximately 4.1 Å, as determined using OVITO (Fig. S1(a)). This distance lies within the cutoff radius (6.29 Å) of the embedded-atom method (EAM) potential [1] employed in this work. However, it is considerably larger than the 1st NN distance (2.86 Å) and is instead comparable to the 2nd NN distance in Al, based on the face-centered cubic (FCC) lattice constant of Al, $a = 4.05$ Å.

To validate the assumption that interaction effects beyond nearest neighbors are negligible, we computed solute-solute interaction energies in a $\Sigma 5(210)$ Al bicrystal using the approach outlined by Nenninger and Sansoz [2]. Our findings reveal that the average Ni-Ni interaction energy between an interstitial Ni atom and its 1st NNs, approximately 7.6 kJ/mol, is notably greater than that between two neighboring interstitial Ni atoms, around 4.3 kJ/mol (Fig. S 1(b)). Furthermore, each interstitial Ni atom is surrounded by 9–11 1st NN substitutes but only two 2nd NN interstitials. This suggests that considering interactions up to the 1st NN distance is sufficient for capturing solute-solute effects, in line with prior studies [3–5]. This simplification improves computational efficiency without significantly affecting segregation predictions. Additionally, the positive interaction energies also reveal repulsion between segregated Ni atoms, which explains their preference for interstitial sites while most nearby substitutional positions remain occupied by Al atoms.

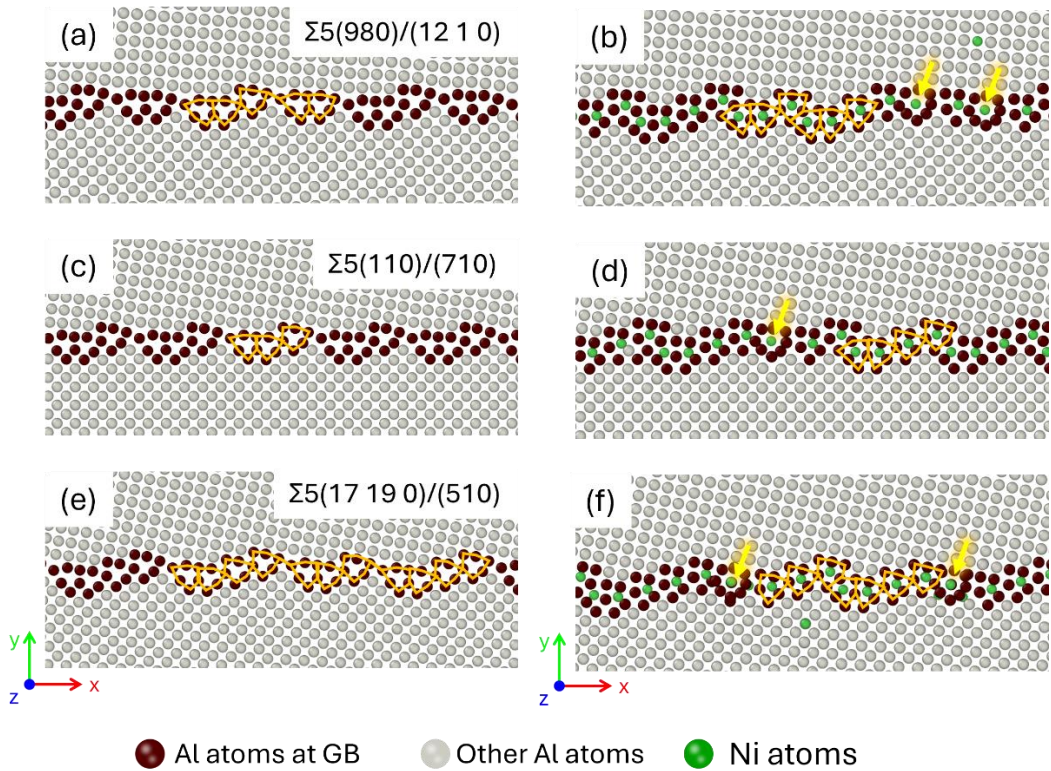

**Fig. S2** GB interstitial segregation in Al Σ5(980)/(12 1 0), Σ5(110)/(710), and Σ5(17 19 0)/(510) asymmetric tilt grain boundaries (ATGBs). Left panels show the original GB structures prior to segregation. Right panels show the corresponding GB structures after hybrid molecular dynamics (MD)/Monte Carlo (MC) simulations at 300 K with a solute concentration of 0.5 at.%. The bright yellow arrows highlight the interplanar segregation Ni atoms.

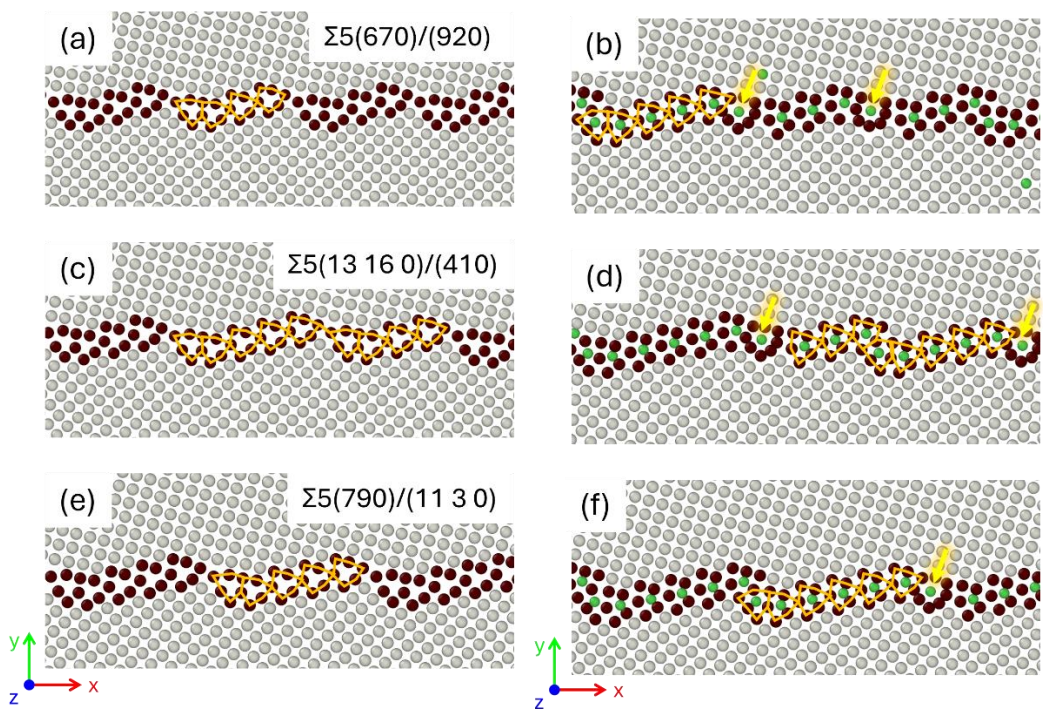

**Fig. S3** GB interstitial segregation in Al Σ5(670)/(92 0), Σ5(13 16 0)/(410), and Σ5(790)/(11 3 0) ATGBs. Left panels show the original GB structures prior to segregation. Right panels show the corresponding GB structures after hybrid MD/MC simulations at 300 K with a solute concentration of 0.5 at.%. The bright yellow arrows highlight the interplanar segregation Ni atoms.

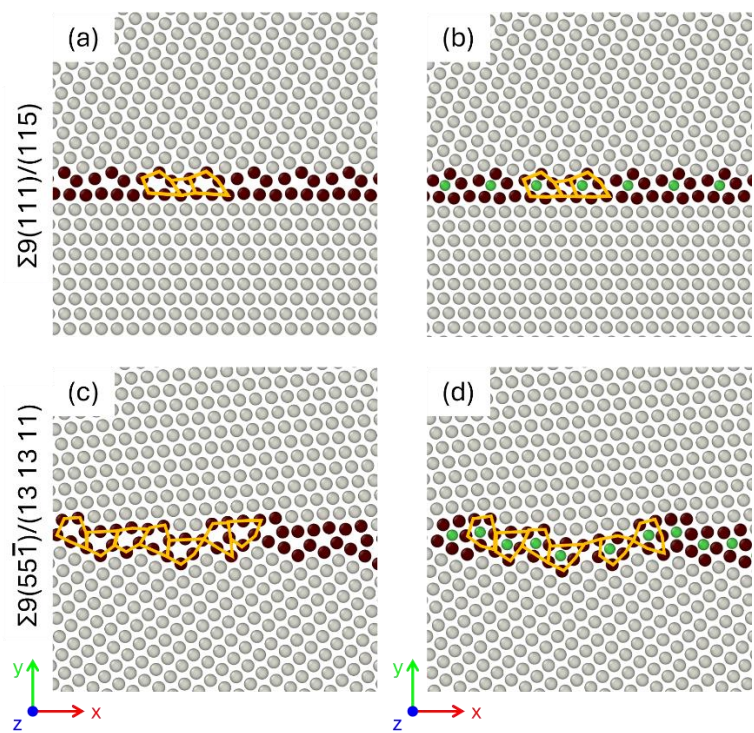

**Fig. S4** GB interstitial segregation in Al $\Sigma 9(111)/(115)$ and $\Sigma 9(55\bar{1})/(13\ 13\ 1)$ ATGBs. Left panels show the original GB structures prior to segregation. Right panels show the corresponding GB structures after hybrid MD/MC simulations at 300 K with a solute concentration of 0.5 at.%.

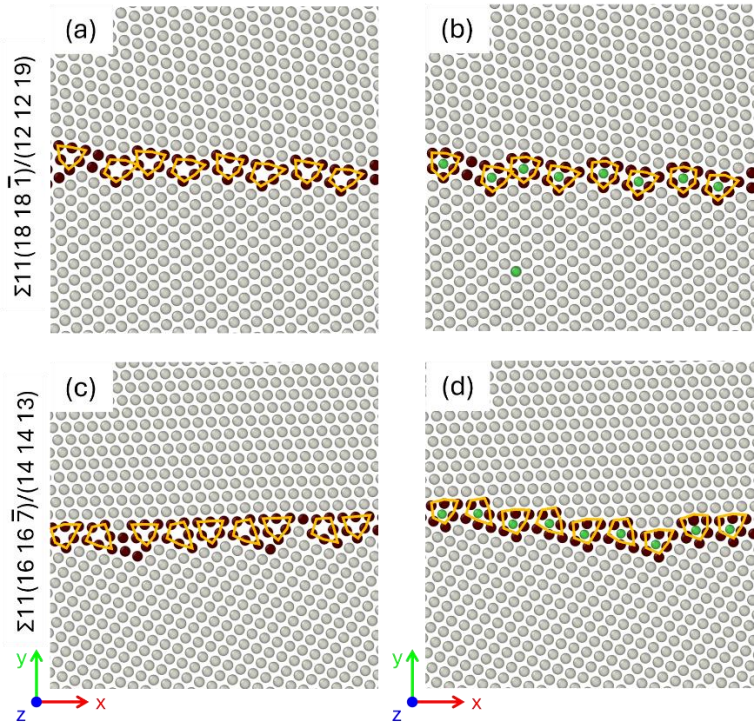

**Fig. S5** GB interstitial segregation in Al $\Sigma 11(18\ 18\ \bar{1})/(12\ 12\ 19)$ and $\Sigma 11(16\ 16\ \bar{7})/(14\ 14\ 13)$ ATGBs. Left panels show the original GB structures prior to segregation. Right panels show the corresponding GB structures after hybrid MD/MC simulations at 300 K with a solute concentration of 0.5 at.%.

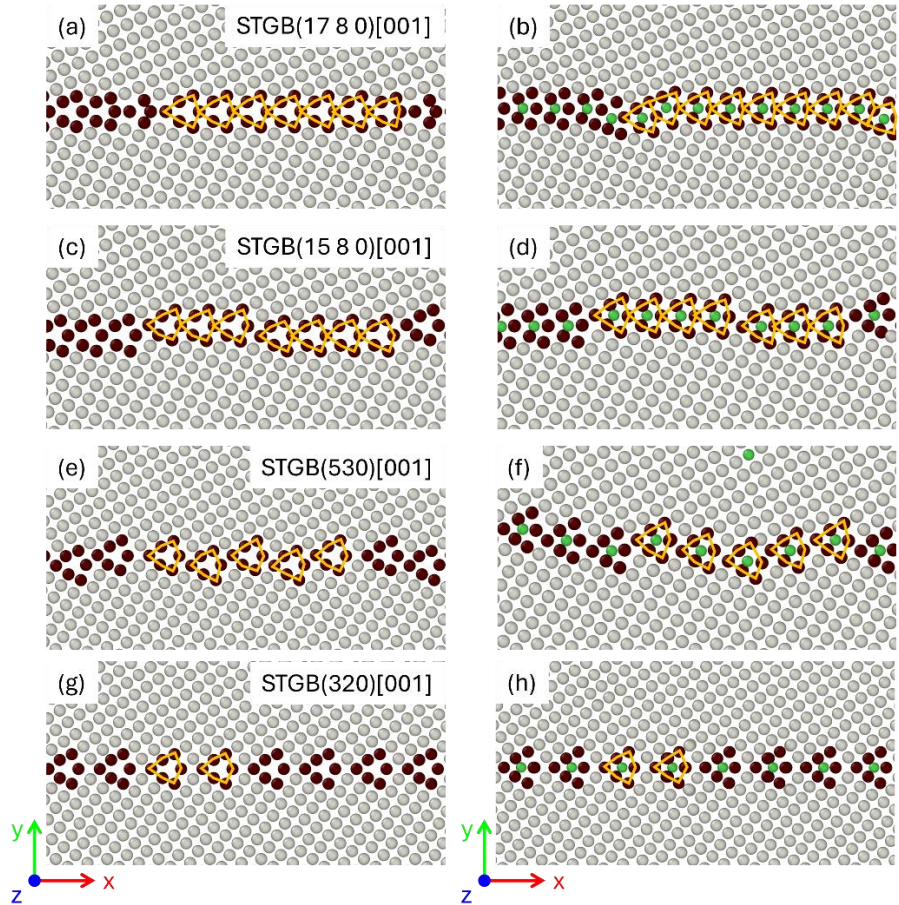

**Fig. S6** GB interstitial segregation in several [001] STGBs of Al. Left panels show the original GB structures prior to segregation. Right panels show the corresponding GB structures after hybrid MD/MC simulations at 300 K with a solute concentration of 0.5 at.%.

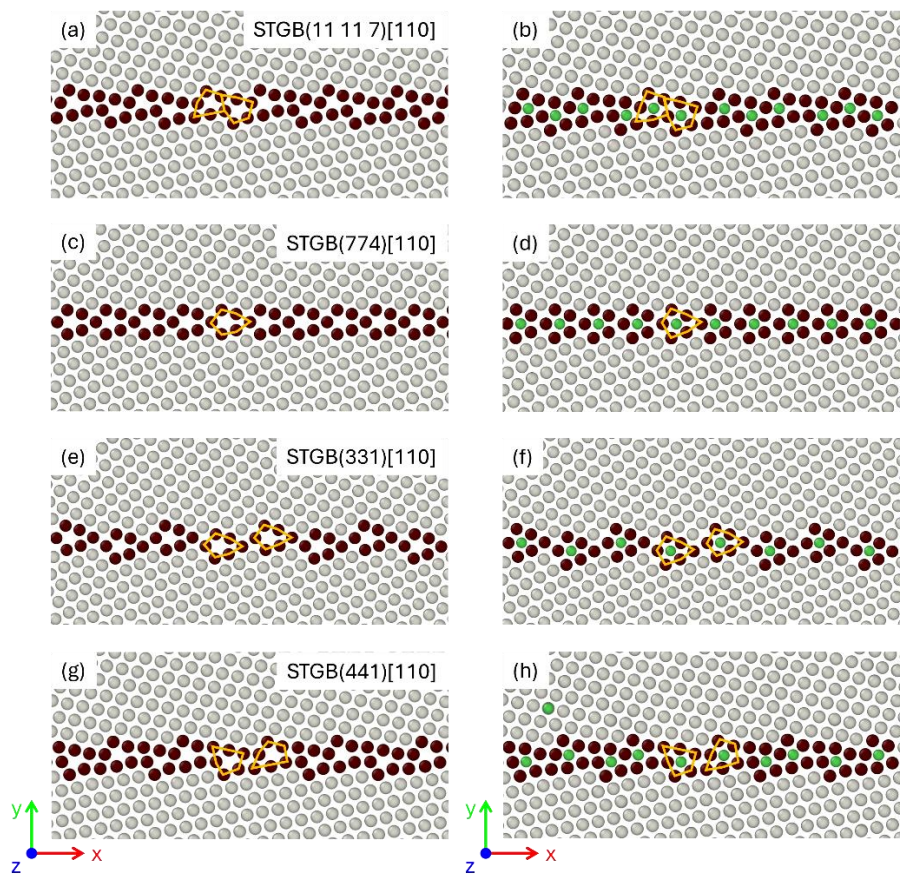

**Fig. S7** GB interstitial segregation in several [110] STGBs of Al. Left panels show the original GB structures prior to segregation. Right panels show the corresponding GB structures after hybrid MD/MC simulations at 300 K with a solute concentration of 0.5 at.%.

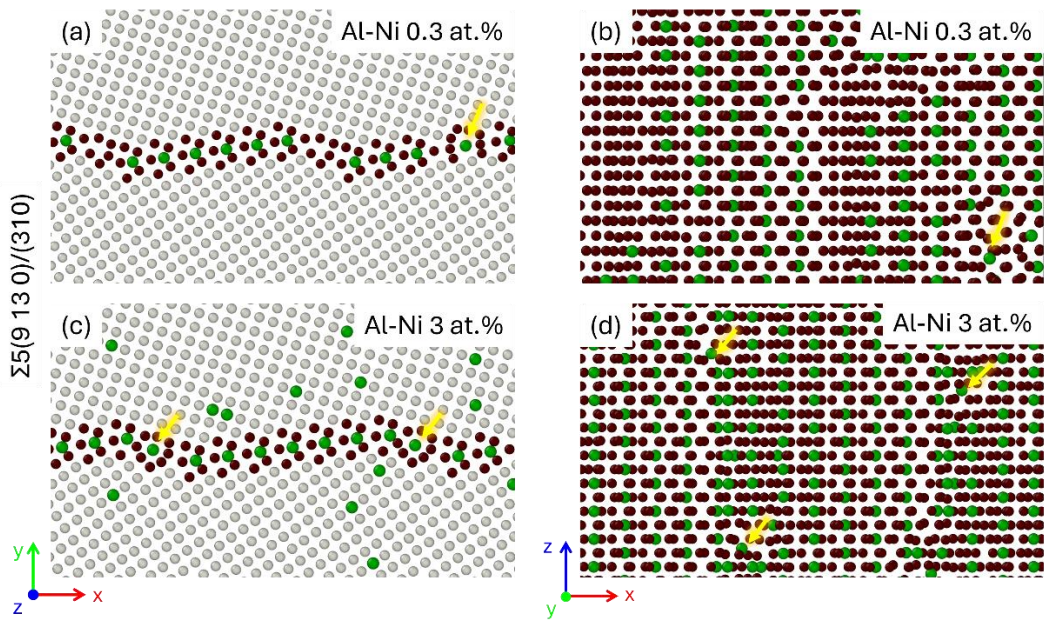

**Fig. S8** Hybrid MD/MC results at 300 K with different solute concentrations for the Σ5(9 13 0)/(310) faceting ATGB. (a) and (b) are sliced top and side views with a solute concentration of 0.3 at.%. (c) and (d) show the sliced top and side views with a solute concentration of 3 at.%. Bright yellow arrows highlight the interplanar interstitial segregation.

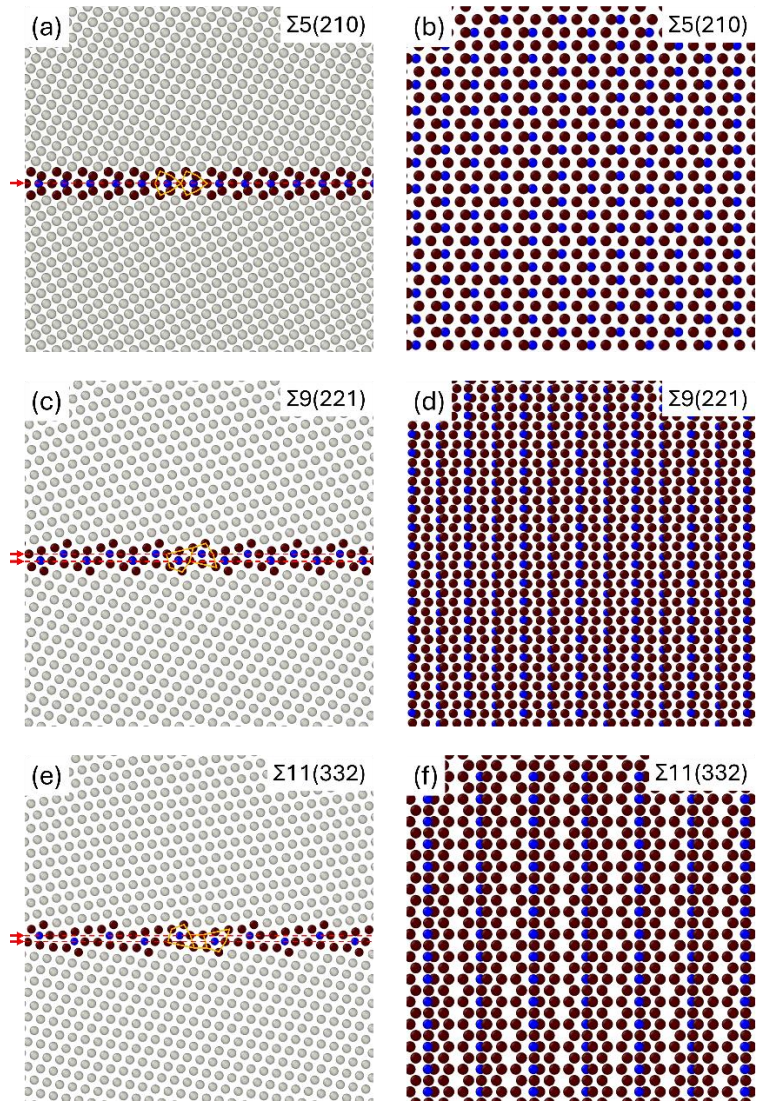

**Fig. S9** Identified interstitial candidate sites within the three Al Σ5(210), Σ9(221), and Σ11(332) STGBs, respectively. The identification results exhibit the same interstitial segregation distribution as the corresponding hybrid MD/MC simulations at 300 K.

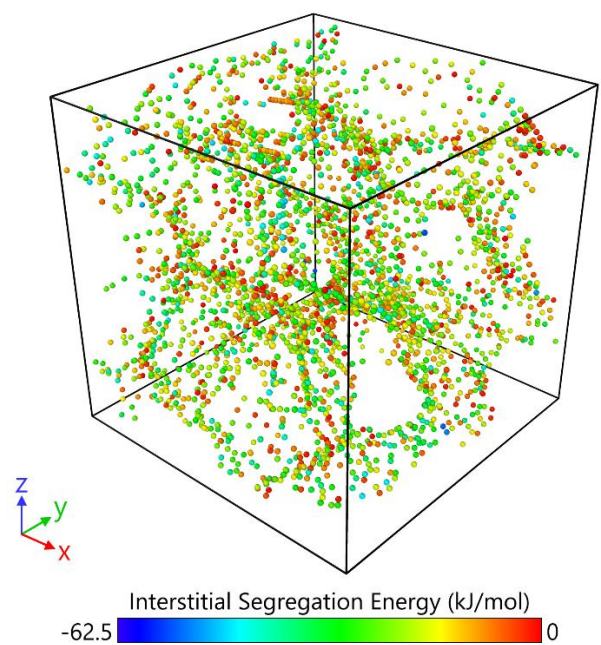

**Fig. S10** Plot of favorable interstitial segregation sites for Ni are well dispersed across the entire GB network of the Al-16 NC sample.

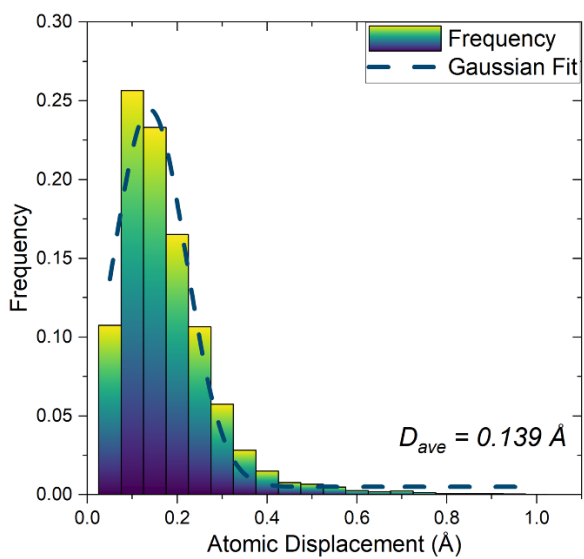

**Fig. S11** Distribution of displacements of the identified interstitial sites during the molecular statics calculations for the substitutional Al-Ni alloy system. The $D_{ave}$ = 0.139 Å refers to the average atomic displacement.

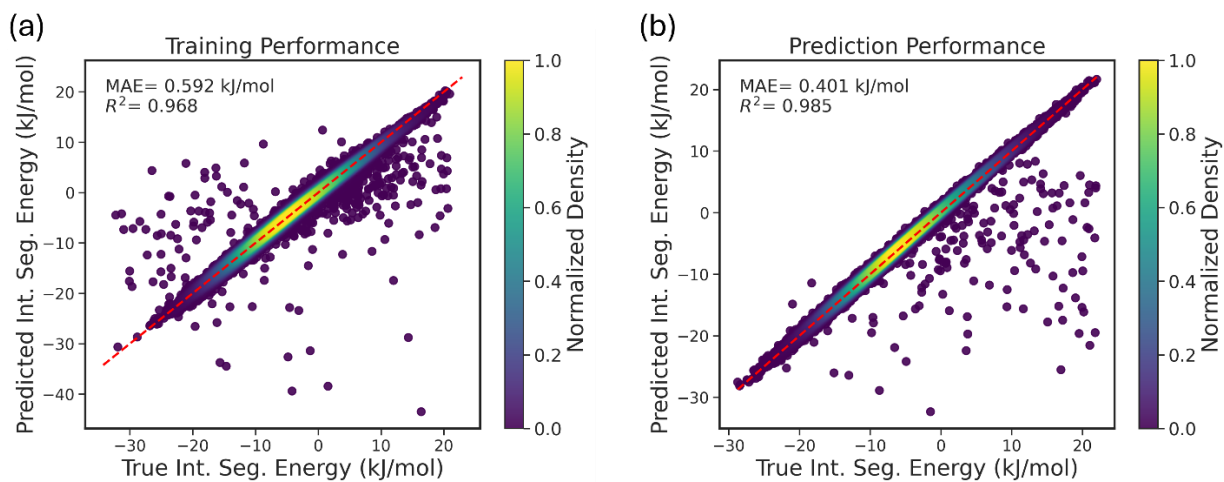

**Fig. S12** (a) The training performance of interstitial segregation energy for the Pd-H system ($16^3$ $nm^3$) using the linear regression method. (b) Prediction performance of the interstitial segregation energy for a large ($20^3$ $nm^3$) nanocrystalline Pd-H structure using the trained linear model.

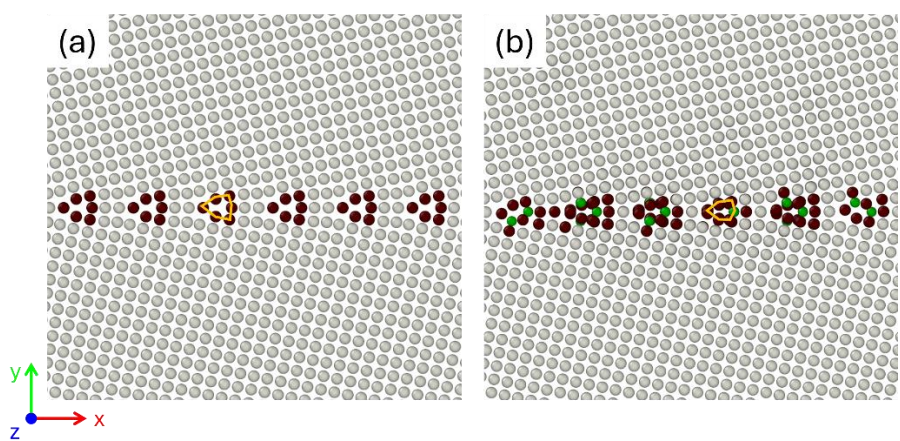

**Fig. S13** (a) Original GB structure of the $\Sigma 13(510)$ STGB. (b) GB structure after hybrid MD/MC simulations at 300 K with a solute concentration of 0.5 at.% for the Al-Ni system.

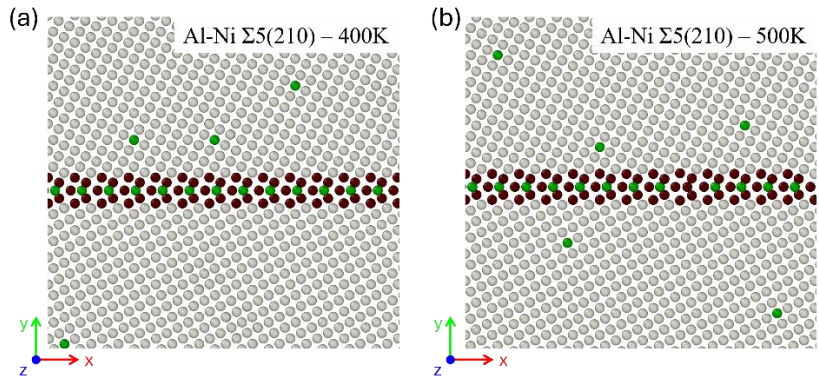

**Fig. S14** Results of hybrid simulations at 300 K in the Al-Ni system at (a) 400 K and (b) 500 K, respectively, with the target solute concentration of 1 at.%. Apparently, interstitial segregation of Ni atoms in Al Σ5(210) STGB occurred in both cases.

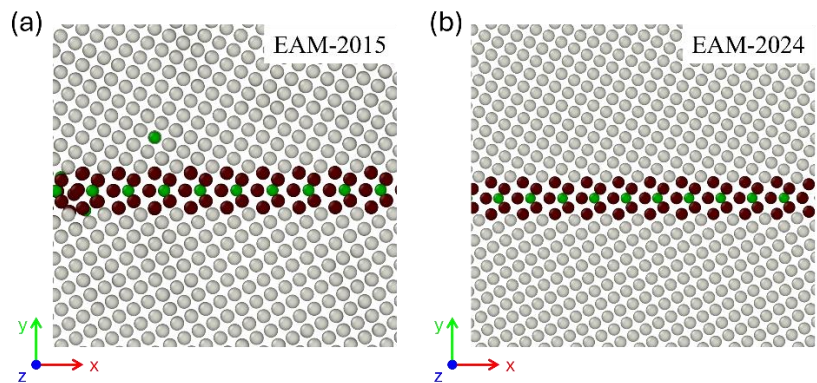

**Fig. S15** Hybrid simulation results using two additional EAM potentials [6,7] showed that interstitial segregation was also clearly observed in both cases. This further confirms the reliability of the findings in this study obtained after hybrid simulations using the interatomic potential for the Al-Ni system [1].

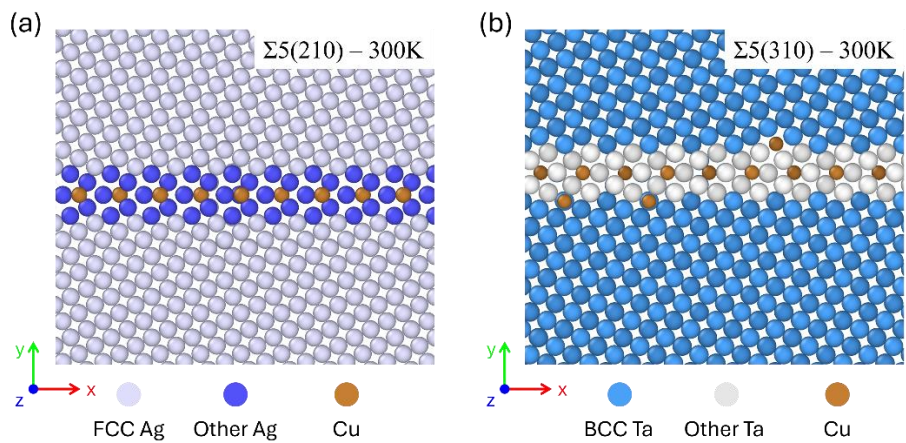

**Fig. S16** Hybrid simulation results for the FCC Ag-Cu and BCC Ta-Cu systems showed that Cu atoms prefer segregate to interstitial sites within the corresponding STGBs of Ag and Ta.

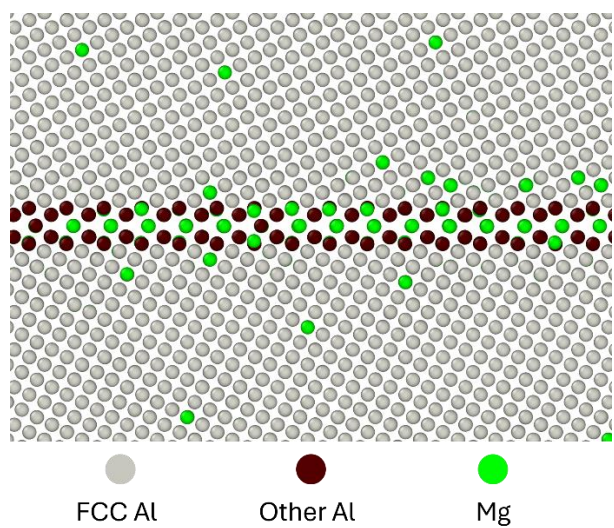

**Fig. S17** Hybrid simulation results for the Al-Mg system at 300 K with the global Mg content of 2 at.% employing the EAM interatomic potential developed by Liu and Adams [8].